\documentclass[english,a4paper,11pt]{article}
\usepackage[margin=3cm]{geometry}
\usepackage[latin1]{inputenc}
\usepackage{latexsym}
\usepackage{booktabs}
\usepackage{amsmath}
\usepackage{textcomp}
\usepackage{algorithm}
\usepackage{makecell}
\usepackage{latexsym}
\usepackage{pdflscape}
\usepackage[final]{graphicx}
\DeclareGraphicsExtensions{.jpg,.jpeg,.pdf,.png,.mps}
\usepackage{epsfig}
\usepackage[round]{natbib}
\usepackage{rotating}
\usepackage{color}
\usepackage{graphicx,subfig}
\usepackage{colortbl}
\usepackage{float}

\usepackage{csquotes}
\usepackage[misc]{ifsym}
\usepackage{multirow}
\usepackage[toc,page]{appendix}
 \usepackage{threeparttable}
\usepackage[bitstream-charter]{mathdesign}
\usepackage[T1]{fontenc}
\usepackage{lmodern}
\usepackage{xcolor}
\usepackage[
    colorlinks=true,
    citecolor=blue,
    linkcolor=blue,
    urlcolor=blue
]{hyperref}

\title{Carbon cost pass-through rate in power system: evidence from Italy under the EU ETS}

\author{$\mathrm{Pierdomenico \ Duttilo}^\mathrm{*,\hspace{0.5mm}\textrm{\Letter}},  
	\  \mathrm{Francesco \ Lisi}^\mathrm{*}$
	\\  
	$^\mathrm{*}$\small{\emph{Department  of Statistical Sciences,  University of Padua, Italy}}\\
    $^\mathrm{\textrm{\Letter}}$\small{\emph{Corresponding author: \href{mailto:pierdomenico.duttilo@unipd.it}{\textcolor{blue}{pierdomenico.duttilo@unipd.it}}}}
}
\date{}

\begin{document}
	\maketitle
\begin{abstract}
This paper investigates the impact of carbon pricing under the EU Emissions Trading System (EU ETS) on the Italian electricity market, focusing on the carbon cost pass-through rate (CPTR) across market zones during Phases~3 and 4 (2016-2024). Using daily data, the study applies an econometric framework based on a linear regression model with autoregressive dynamics to estimate the extent to which carbon costs are reflected in wholesale electricity prices. It further incorporates robustness checks and quantile regression to assess how the CPTR varies across different fuel spread levels. The results show that carbon costs are positively and significantly transmitted to electricity prices, confirming the relevance of carbon pricing as a key market driver. However, pass-through is incomplete, with CPTR values consistently below $100\%$. At the national level, the pass-through estimate is around 32\%, with no statistically significant change between Phase~3 and Phase~4. Substantial heterogeneity emerges across market zones: pass-through increases in the North, Centre-North, and Sardinia during Phase 4, while it declines in the Centre-South and Sicily, reflecting differences in generation mix, carbon intensity, and market conditions. Overall, the findings highlight the importance of market zones factors in shaping the effectiveness of carbon pricing in electricity markets.
\noindent 
\hspace{1cm}
\end{abstract}

\section{Introduction}\label{sec.intro}
Pricing carbon dioxide (CO$_2$) emissions, through a carbon tax or an emissions trading system (ETS), represents a cost-efficient way to reduce CO$_2$ emissions from carbon-intensive sectors and mitigate the adverse impacts of climate change \citep{baumol1988theory,Metcalf2009,stiglitz2017report,Geroe2019,STIGLITZ2019594, rafaty2025carbon}.

The introduction of carbon pricing has strongly affected the power sector, leading to a structural change in the marginal cost of electricity generation. As electricity is commonly traded on wholesale markets, market prices can incorporate the costs of emitting CO$_2$ \citep{NAZIFI2021}. In this context, the carbon cost pass-through rate (CPTR) is the extent to which carbon costs are passed through the electricity market prices \citep{SIJM2006,SIJM2008,CHEN2008,LISE2010,LIN2019,ZHU2019202,NAZIFI2021}. The CPTR quantifies how much of the carbon costs incurred during electricity production are transferred to consumers through higher electricity prices \citep{DING2022}. It is also defined by \cite{SIJM2006} and \cite{CHEN2008} as ``the ratio of changes in the price of power to the changes in marginal costs due to the implementation of the carbon price mechanism'' \citep{NAZIFI2016}. A higher CPTR means that consumers pay most of the cost of the pricing mechanism, while a lower CPTR indicates that generators absorb a larger share. It may also suggest a shift in the marginal cost structure of electricity generation among different fuel types \citep{NAZIFI2016,NAZIFI2021}. According to economic theory, the CPTR depends on the method used to allocate carbon emission allowances (auctioned or freely allocated) and, to a large extent, on the structure of the power market \citep{SIJM2008}. The structure of the power market is determined by the interaction of three key factors \citep{SIJM2012}: the number of firms operating in the market (market concentration or competitiveness), the shape of the demand curve (linear or iso-elastic), and the behaviour of marginal costs (constant or variable). Thus, estimating CPTR offers valuable insight for market participants, regulators, and climate policy-makers on the effectiveness of carbon pricing policies and their economic impact.

The first carbon taxes were adopted by Finland and Poland in 1990, while the world's first ETS was launched by the European Union (EU) in January 2005 as a policy instrument to reduce GHG emissions under the Kyoto Protocol \citep{Parker2006}. Since then, the number of ETSs has increased to 36, with the latest additions from the major economies such as China, Australia, and Japan \citep{rafaty2025carbon}.\\
The EU ETS has evolved through four distinct phases. Phase 1 (2005-2007) was a pilot that tested the system and set up reporting infrastructures. The power sector has been included in the EU ETS since Phase 1 covering heavy energy-intensive installations consisting of power plants and other combustion stations with a thermal input of $\geq$20MWh \citep{EC_ETSHandbook2015}. The aim was to encourage investment in more sustainable energy sources and to provide an incentive to Europe's largest emitters to improve energy efficiency.\\ Phase 2 (2008-2012) aligned with the Kyoto Protocol and expanded geographic and sectoral coverage. During Phases 1 and 2, allowances were mostly allocated for free based on national allocation plans (NAPs).\\ Phase~3 (2013-2020) introduced a centralised EU-wide cap, shifted to auctioning as the main allocation method, and applied a linear reduction factor (LRF) to achieve significant emission cuts. The allocation of allowances for the power sector was 100\% auctioned or sold \citep{EC_ETSHandbook2015}.\\ Phase~4 (2021-2030) raises the climate ambitions under the European Climate Law \citep{EuropeanClimateLaw2021} by increasing the LRF and expanding the sectoral coverage. 

\begin{figure}[H]
    \centering
    \includegraphics[width=0.9\textwidth]{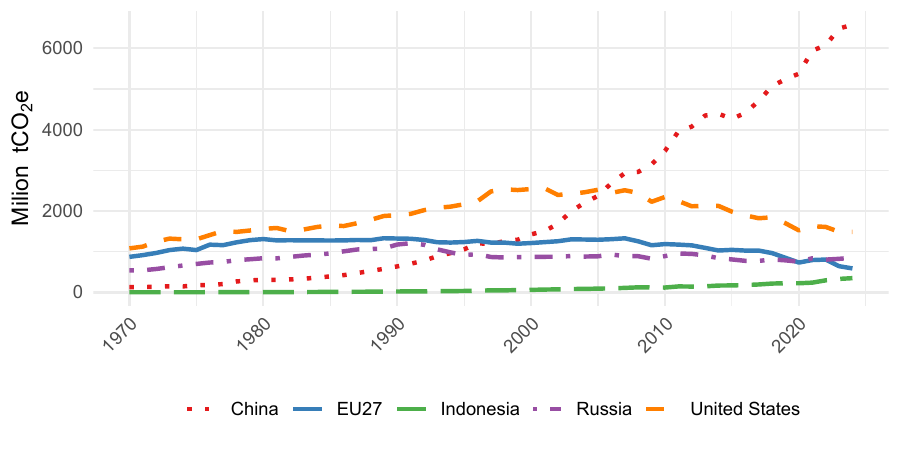}
    \caption{CO$_2$ emissions (in million tCO$_2$e) in the power industry of top emitting economies, 1970-2024.}
    \label{fig:co2_by_topc}
\end{figure}

The purpose of EU ETS is to make fossil fuel combustion more expensive, encouraging electricity producers to reduce coal-based generation and to move to gas and renewable energy sources \citep{NAZIFI2021,DING2022}.\\ In 2005, the European power sector was the third-largest carbon emitter, after China and the United States (Figure \ref{fig:co2_by_topc}), accounting for 35.10\% of total emissions in the EU. Among EU countries, Germany exhibited a higher proportion at 41.7\%, indicating a strong dependence on carbon-intensive electricity production, while Italy's power sector contributed 32.2\% of total CO$_2$ emissions of the country, slightly below the EU27 average \citep{Crippa2025}.\\ The EU Member States reported a 15.5\% reduction in CO$_2$ emissions in 2023 compared to 2022, bringing EU ETS emissions to approximately 47\% below 2005 levels \cite{EUco22024}. This highlights the effectiveness of the EU's cap-and-trade system as a key driver of decarbonisation. The decrease was primarily driven by the power sector, where emissions fell by 24\% due to a substantial increase in renewable electricity generation (wind and solar) and the continued substitution of coal for gas. Furthermore, the European power sector dropped to fourth place, after Russia, the United States and China (Figure \ref{fig:co2_by_topc}), accounting for 23.9\% of total emissions in the EU.
Among EU Member States, Germany recorded the highest share at 29.6\%, while Italy's power sector contributed 23.1\% of the country's total CO$_2$ emissions \citep{Crippa2025}.

This study aims to examine the impact of the EU ETS on the Italian electricity market. Although part of the existing literature \citep{Levy2005,Chernyavs2008,CHERNYAVSKA2008,SIJM2008,JOUVET20131370} focusses on the effects of EU ETS before and after its introduction (Phases 1 and 2), the impacts of Phases 3 and 4 have not yet been fully investigated. The latter are characterised by a progressive tightening of emission caps, which may have significantly influenced electricity prices and overall carbon emissions. In addition, both phases coincide with major global shocks arising from the COVID-19 pandemic and the Russia-Ukraine war, which contribute to the change of the energy fuel mix \citep{Ghiani2020,MALISZE2024,Liu04052025,Emiliozzi2025}. The analysis addresses the following research questions:
\begin{itemize}
\item RQ1: How do the Italian electricity market zones differ in their responses to carbon price fluctuations?
\item RQ2: How does the carbon cost pass-through rate vary between EU ETS Phase 3 and Phase 4?
\item RQ3: Does the carbon price transmission rate vary across different levels of fuel spreads?
\end{itemize}

To explore these questions, the empirical framework developed by \cite{SIJM2008}, \cite{CHEN2008}, \cite{NAZIFI2016}, \cite{NAZIFI2021}, and \cite{DING2022} is adopted, in which a linear regression model is used to estimate the CPTR. The baseline specification is augmented to address stationarity and autocorrelation in the data. Alternative model specifications are also considered to assess the robustness and reliability of the results. In addition, quantile regression is employed to examine the behaviour of the CPTR across different levels of the spread between wholesale electricity spot prices and fuel costs, rather than focusing solely on its mean effect.\\
This paper contributes to the literature in several ways. First, it provides updated empirical evidence on carbon cost pass-through in the Italian electricity market during the latest phases of the EU ETS, considering both the aggregate level and individual market zones, and highlighting zonal heterogeneity. Second, it contributes methodologically by extending the baseline specification to address stationarity and autocorrelation in the data. In addition, the analysis goes beyond average effects by exploiting quantile regression, allowing the CPTR to vary across the spread distribution.

The remainder of the paper is structured as follows. Section~\ref{sec.background} provides background on the evolution of the EU ETS, the Italian electricity market, and the related literature. Section~\ref{sec.data} describes the data. Section~\ref{sec.methodology} outlines the empirical methodology used to estimate the CPTR. Section~\ref{sec.results} presents the empirical results and robustness checks. Finally, Section~\ref{sec.conclusions} concludes with a discussion of the main findings and policy implications.

\section{Background}\label{sec.background}

\subsection{Evolution of the EU ETS}\label{sec.euets}
The EU ETS is a ``cap-and-trade'' system that limits the total amount of GHG emitted by regulated entities. The cap determines the number of allowances available. Each allowance represents the right to emit one tonne of CO$_2$ equivalent (tCO$_2$e). Participants must surrender allowances equal to their verified emissions, while surplus allowances can be traded, ensuring that emission reductions occur where they are most cost-effective \citep{EC_ETSHandbook2015}. The evolution of the EU ETS has been structured into four trading phases.

\textbf{Phase 1 (2005-2007)} was the pilot phase of the EU ETS. Its central objective was to validate the base elements of a carbon market by testing the mechanisms of price formation and establishing the infrastructure to account for and report CO$_2$ emissions. The system initially included 25 Member States, expanding to 27 in 2007. The industrial coverage encompassed CO$_2$ emissions from large fixed installations, notably power plants, oil refineries, and key manufacturing processes. The total emission cap was calculated bottom-up, based on the aggregation of the NAPs of each Member State \citep{EC_ETSHandbook2015}. Allowances were predominately allocated free of charge to participants, determined through Member State-specific NAPs. 

\textbf{Phase 2 (2008-2012)} was structurally aligned with the first commitment period of the Kyoto Protocol. Geographically, the system covered the entire European Economic Area (EEA) and also Norway, Iceland, and Liechtenstein. Allocation continued primarily through free distribution through the NAPs. The sectoral scope was notably expanded to include the aviation sector at the beginning of 2012 \citep{EC_ETSHandbook2015}. 

\textbf{Phase~3 (2013-2020)} established a more centralised and market-driven scheme. The core goal was to address the inefficiencies and competitive distortions that arise from the NAP system of the previous two phases. The system transitioned from NAPs to a single, unified EU-wide cap. This cap was fixed at 2,084 million allowances in 2013, subject to an annual linear reduction factor (LRF) of 1.74\% (38 million tCO$_2$ per year) relative to 2010 levels \citep{EC_ETSHandbook2015,EC_EU_ETS_Emissions_Cap}. This mechanism guaranteed a cumulative emission reduction target of 21\% compared to 2005 levels by 2020. Auctioning became the default allocation method, projected to account for approximately 50\% of total allowances. The power generation sector was subject to 100\% auctioning from 2013, with only specific exceptions in eligible Member States. The sectoral coverage also included CO$_2$ emissions from aluminium and chemical production processes.

\textbf{Phase~4 (2021-2030)} represents a fundamental intensification of the EU's flagship climate policy. It plays a key role in the implementation of the European Climate Law, which commits the EU to achieve climate neutrality by 2050. In addition, as an intermediate objective, the EU aims to reduce net greenhouse gas (GHG) emissions by at least 55\% by 2030 compared to 1990 levels \citep{EuropeanClimateLaw2021,EUETS2025,DIRCONSEUETS2025}. The annual LRF was initially increased from 1.74\% to 2.2\% starting in 2021. Following the subsequent revision in 2023, the LRF was further amplified to 4.3\% for the period 2024 to 2027 and 4.4\% from 2028 onwards. This increased ambition sets the overall cap to reduce the emissions of the covered sector' by 62\% compared to 2005 levels by 2030 \citep{EC_EU_ETS_Emissions_Cap}. Furthermore, the cap underwent explicit downward adjustments in two stages: a reduction of 90 million allowances in 2024 and an additional 27 million allowances in 2026. From 2024, the scope was extended to cover CO$_2$ emissions from large ships operating in EU ports and additional flights. Phase~4 implements key structural changes to the allocation of allowances. Free allocation to high-risk industrial sectors (such as iron, steel, cement, aluminium, fertilisers, and hydrogen) will gradually be phased out between 2026 and 2034. The carbon border adjustment mechanism replaces free allowances as the main anti-carbon leakage measure. From 2026 onwards, free allocations are conditioned to the adoption of energy efficiency improvements or decarbonisation plans.

\begin{figure}[H]
    \centering
    \includegraphics[width=0.9\textwidth]{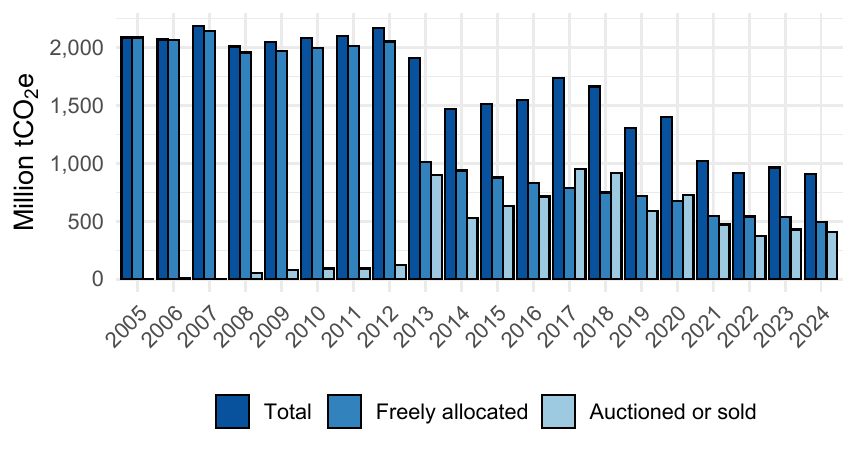}
    \caption{EU ETS allowances (in million tCO$_2$e) by year.}
    \label{fig:allow_by_years}
\end{figure}

Figure \ref{fig:allow_by_years} reports the total volume of EU allowances, distinguishing between freely allocated and auctioned or sold allowances, for each year \citep{EEA_ETS_data}. The figure highlights the two important developments in the evolution of the EU ETS: a gradual reduction in the overall cap and a shift in the allocation mechanism from free allocation to auctioning. During the first two phases (2005-2012), the cap volume remained relatively high and stable, fluctuating around 2,095 million tCO$_2$e. From 2013 onwards, the total cap exhibits a clear and sustained year-over-year decline, reflecting the implementation of a linear reduction factor. The first years of Phase~4 (2021-2024) show a continued rapid decrease, reaching the lowest observed level of 905,774,912 tCO$_2$e. EU allowances decreased by 56.7\% between 2005 and 2024.

\subsection{The Italian electricity market}\label{sec.itamarket}
The Italian power exchange is managed by the Gestore del Mercato Elettrico (GME) and consists of several markets depending on the type of electricity traded and the delivery timing \citep{GME_2025}. This analysis focusses on the day-ahead market\footnote{The Italian acronym is Mercato del Giorno Prima (MGP).}, where producers, wholesalers, and consumers trade electricity for each hour of the following day. In the day-ahead market, generators submit offers based on the output of individual plants. Prices are determined through uniform-price auctions, setting the system marginal price for each hour. Generators whose offers are accepted receive this price for the electricity supplied in their zone, while consumers pay a weighted average of zonal prices according to the volume of electricity exchanged \citep{Caporin2022}.

The day-ahead market is divided into seven market zones that share the same pricing rules and market design North, Centre-North, Centre-South, South, Calabria, Sicily, and Sardinia. Although producers in each zone have access to the same primary energy markets, such as natural gas and coal, the zones differ in terms of electricity demand and supply structures. Table \ref{tab.energy_gen_by_source} shows important differences in the energy mix of the seven market zones \citep{Terna_Generazione_2025}. In 2024, thermal power dominates throughout the country, but ranges from about 46\% in the Centre-North to over 70\% in Calabria. It should be noted that the number of thermal power plants (N.Thermal) is highest in the North, with 3,787 units, while smaller market zones such as Sardinia and Calabria have considerably fewer thermal plants, with 48 and 69 units, respectively \citep{Terna2024_GenPlants}. The North relies heavily on hydroelectric generation, while the Centre-North benefits from significant geothermal capacity. The southern zones and the islands show higher shares of wind and photovoltaic (PV) generation, particularly in the South and Sicily.

The seven market zones of the Italian electricity market provide an interesting case study to examine differences in CPTR within a single country operating under the EU ETS \citep{Chernyavs2008}.

\begin{table}[ht]
\centering
\begin{tabular}{lcccccc}
\toprule
Zone         & Thermal & Geothermal & Hydro   & Wind & PV   & N.Thermal\\
\midrule
Italy        & 55.6    & 2.1        & 20.5    & 8.4  & 13.4 &  5,0005\\
North        & 55.6    & 0.1        & 33.5    & 0.3  & 10.5 &  3,787 \\
Centre-North & 46.2    & 29.6       & 6.2     & 1.5  & 16.5 &  339   \\
Centre-South & 54.3    & 0.0        & 12.0    & 13.2 & 20.5 &  500   \\
South        & 52.4    & 0.0        & 1.1     & 29.3 & 17.2 &  164   \\
Calabria     & 70.2    & 0.0        & 6.2     & 16.7 & 6.9  &  69    \\
Sicily       & 55.0    & 0.0        & 1.5     & 24.5 & 19.0 &  98    \\
Sardinia     & 66.2    & 0.0        & 3.2     & 15.0 & 15.6 &  48    \\
\bottomrule
\end{tabular}
\caption{Percentage of energy generated in each zone by energy source in 2024. PV stands for photovoltaic.}
\label{tab.energy_gen_by_source}
\end{table}

\subsection{Literature review}\label{sec.literature}
The empirical literature on CPTR is extensive and can be classified into two main categories: ex-ante projections \citep{Linares2006,KARA2008193,LISE2010} and ex-post evaluations based on observational data \citep{rafaty2025carbon}. Within the latter category, numerous studies have examined the extent to which carbon emission costs under the EU ETS are passed through to final consumers \citep{Levy2005,SIJM2006,SIJM2008,Chernyavs2008,CHERNYAVSKA2008,Fell2010,GRONWALD2011,SIJM2012,Fabra2014,HUISMAN2015,JOUVET20131370,Mazzarella2017,JI2019,ZHU2019202,Caporin2021}. Most of these studies focus on the first two phases of the EU ETS \citep{Levy2005,SIJM2006,SIJM2008,Chernyavs2008,CHERNYAVSKA2008,Fell2010,GRONWALD2011,SIJM2012,JOUVET20131370}. Empirical studies highlight significant differences in CPTRs among EU member states.\cite{SIJM2006} estimated CPTRs using a regression model, finding values ranging 60-117\% for Germany and 64-81\% for the Netherlands. \cite{SIJM2008} extended the analysis to additional European markets and reported a positive but incomplete CPTR in the Netherlands, Germany, France, and Sweden, while identifying a complete CPTR in the United Kingdom. \cite{JOUVET20131370} investigated several European countries during Phases 1 and 2, showing that CPTRs decreased in Phase 2 and became negative in some markets. Similar results were found by \cite{Ahamada2015, Ahamada2018}. \cite{Fabra2014} analysed the Spanish electricity market and find that the CPTR varies by hours, ranging 70-140\% during peak hours and 28-97\% during off-peak periods, depending on the model specification.

For Italy, the main findings of the literature are summarised in Table~\ref{tab.cptr_italy_review}. The evidence on CPTR in the Italian electricity market appears highly heterogeneous, both in terms of data and methodological approaches. \cite{Levy2005} analysed the relationship between CO$_2$ and wholesale electricity prices in 2005 using linear regression models. The results showed that the CPTRs shifted from negative to positive values between the first and second quarters of 2005, indicating unstable cost pass-through during the initial phase of the EU ETS. \cite{Chernyavs2008} used a different approach based on the simulation of the load and merit-order supply curves for the entire Italian electricity market, as well as for the North and South zones. The results showed that the CPTR depends on peak and off-peak hours and market competitiveness. Similarly, \cite{CHERNYAVSKA2008} estimated the CPTR for the Italian market in 2005 and 2006, finding that it changed from 0 in 2005 to a range between -50\% and 20\% in 2006. Using the regression model, \cite{SIJM2008} found negative coefficients during the first Phase. \cite{JOUVET20131370} showed that CPTR decreased or even became negative during Phase 2. In contrast, \cite{Mazzarella2017} identified a progressive increase in cost transmission over time, with CPTR estimates rising from near zero in 2005-2007 to 71\% in 2008-2012 and 82\% in 2013-2015, suggesting a more efficient reflection of carbon costs in electricity prices as the EU-ETS matured. \cite{Caporin2021}, used a vector error correction model (VECM) to investigate the relationship between the EU-ETS allowance prices and the Italian wholesale electricity price during Phase~3 (2013-2018). The analysis estimated the long-term relationships among electricity, natural gas, and allowance prices, distinguishing between peak and off-peak hours. The results indicate a high and increasing long-run pass-through from natural gas to electricity prices, 70\%, while the CPTR remains low, around 7\%. 

As highlighted by the literature, most studies focus on the initial phases of the EU-ETS. Moreover, only \cite{Chernyavs2008} estimated the CPTR for two Italian market zones. The present study aims to address the three research questions (RQ1-3) introduced in Section~\ref{sec.intro}, with the objective of identifying potential gaps between the Italian electricity market zones during the last two phases of the EU-ETS.

\begin{table}[ht!]
\centering
    \resizebox{\linewidth}{!}{
\begin{tabular}{lccccccc}
\toprule
 &  &  &  &  & \multicolumn{3}{c}{\textbf{CPTR}} \\
\cmidrule(lr){6-8}
\textbf{Study} & \textbf{Methodology} & \textbf{Zone} & \textbf{Market} & \textbf{Period} & \textbf{All} & \textbf{Peak} & \textbf{Off-peak} \\
\midrule
\cite{Levy2005} &  Regression model &  Italy  &  Spot  &  Jan.-Mar. 2005  & -1.82 & - & - \\
&   &   &    &  Apr.-Jun. 2005  & 2.04 & - & - \\
\cite{Chernyavs2008} & \makecell{Load duration \\curve approach} & Italy & Spot & 2006 & - & $[1.10;1.50]$ & $[0.90;1.10]$  \\
&  & North & Spot & 2006 & - & $[1.50;2.10]$ &  $[0.90;1.10]$ \\
&  & South & Spot & 2006 & - & $[-0.10;0.50]$ & $[0.90;1.10]$ \\
\cite{CHERNYAVSKA2008} & \makecell{Load duration \\curve approach} & Italy & Spot & 2005 & 0 & - & -\\
& &  &  & 2006 & $[-0.50; 20]$ & - & -\\
\cite{SIJM2008} & Regression model & Italy & Spot & 2005 & - & -0.97 & 0.39\\
&   &   &    &  2006  & - & -0.67 & -2.98 \\
\cite{JOUVET20131370} & Regression model & Italy & Spot & 2005-2006 & - & $[-0.64;1.05]$ & $[-3.56;-0.03]$ \\
& &  &  & 2008-2010 & - & $[-6.39;-1.23]$ & $[-5.43;-1.01]$\\
\cite{Mazzarella2017} & Regression model & Italy & Spot & 2005-2007 & 0.70 & - & -\\
&&&&2008-2012&  0.30 & - & - \\
&&&&2013-2015& $\approx 0$ & - & - \\
\cite{Caporin2021} & VECM model & Italy & Spot & 2013-2018 & 0.77 & 0.78 & 0.76\\
\bottomrule
\end{tabular}
}
\caption{Summary of key contributions on CPTR for Italy. VECM stands for vector error correction model.}
\label{tab.cptr_italy_review}
\end{table}

\section{Data}\label{sec.data}
This section describes the data construction used for the empirical analysis aimed at estimating the CPTR in the Italian power market over the period 2016--2024. 

The data set is constructed on a daily basis and covers the period from 1 January 2016 to 31 December 2024 for Italy and its bidding zones. Phase~3 spans from 1 January 2016 to 31 December 2020, while Phase~4 covers the period from 1 January 2021 to 31 December 2024. The first three years of Phase~3 are excluded due to data limitations. This sample restriction may influence the estimated pass-through if the excluded early years of Phase~3 had different transmission dynamics. However, previous evidence for Italy over 2013--2018 finds low and declining carbon pass-through \citep{Mazzarella2017,Caporin2021}, which is consistent with the pass-through estimated in this study.

Calabria and South zones are also excluded from the analysis. Calabria was introduced as a new market zone in 2021 \citep{Terna2021_NewZones}, while in the South most of the thermal capacity physically located within the zone was excluded until early 2019 \citep{Caporin2021}. As a result, a consistent comparison between Phase~3 and Phase~4 is not feasible for these zones.\\
The main variables include the volume-weighted wholesale electricity spot price, the fuel cost of power generation, and carbon costs.

The wholesale electricity spot prices for the Italian day-ahead market are obtained from GME at hourly frequency \citep{GME_ZonalPrices}. Volume-weighted average daily spot electricity prices ($p^e_t$) are computed as
\begin{equation}
    p^e_t = \sum_{i=1}^{24} \frac{D_i}{\sum_{i=1}^{24} D_i} p_i,
\end{equation}
where $D_i$ and $p_i$ denote the hourly electricity demand and the hourly spot price, respectively. The hourly electricity demand is collected from Terna \citep{Terna_ZonalDemand}, the Italian transmission system operator (TSO). 

Average daily electricity demand $D_t$ is defined as
\begin{equation}
D_t = \frac{1}{24} \sum_{i=1}^{24} D_i .
\end{equation}
It represents an important explanatory variable, as it influences electricity spot prices by reflecting daily, weekly, and annual consumption patterns \citep{CHEN2008, NAZIFI2016}.

The fuel cost of power generation, $p^f_t$, is computed following the methodology proposed by \cite{EIA_DarkSpreads} and applied by \cite{DING2022}. Three main fuels are considered $(j = 1, \ldots, 3)$: coal, oil, and natural gas. Daily fuel prices $(p_{t,j})$ are obtained from \cite{FactSet}: Brent crude oil (\texteuro{}/bbl), Newcastle Coal Near Term (\texteuro{}/t) and Dutch TTF Gas Monthly (\texteuro{}/MWh). The prices of oil and coal are converted into \texteuro{}/MWh using the heat content coefficients provided by \cite{EIA_UnitsCalculators}. The fuel cost of power generation is then calculated as follows:
\begin{equation}
    p^f_t = \frac{\sum_{j=1}^3 p_{t,j} G_{t,j} \eta_j}{\sum_{j=1}^3 G_{t,j}},
\end{equation}
where $\eta_j$ denotes the heat rate associated with each fuel $j$, measured as the energy input required to produce one unit of electricity output. Heat rates are calculated as annual averages over a six-year period (2016--2021), using country-specific data for Italy provided by \cite{ISPRA_Inventory}. The variable $G_{t,j}$ represents the daily electricity generation for each fuel type \citep{Terna_ZonalDemand}.

A key quantity is $s_t = p_t^e - p_t^f$, which represents the spread between the volume-weighted wholesale electricity spot price $p_t^e$ and the fuel cost of electricity generation $p_t^f$.

The daily carbon costs are obtained in two steps. First, daily carbon emissions from electricity generation, denoted as $E_t$, are calculated using the following emission factor-based method \citep{BERTOLINI2025}
\begin{equation}
\begin{split}
    &E_t = \sum_{j=1}^{3} E_{t,j},\\
    &E_{t,j} = G_{t,j} \times EF_j \times O_j \times M,
\end{split}
\end{equation}
where: $E_{t,j}$ denotes the daily carbon emissions for the fuel type $j$, $EF_f$ is the country-specific emission factor for the fuel type $j$;  $O_f$ is the oxidation rate for the fuel type $j$; $M$ is the ratio of the molecular weight of CO$_2$ to the atomic weight of carbon fixed to $44/12=3.6667$.\\
Second, the daily carbon intensity $I_t$ (\text{tCO$_2$e}/\text{MWh}) is defined as the ratio
\begin{equation}
    I_t = \frac{E_t}{\sum_{j=1}^{3}G_{t,j}}.
\end{equation}
The carbon cost $C_t$, i.e., the cost of carbon permits required to cover the emissions from generating one MWh of electricity, is calculated by multiplying the daily EU ETS carbon spot price $p_t^{CO_2}$ (\texteuro{}/tCO$_2$e) by the zonal's carbon intensity \citep{NAZIFI2016,NAZIFI2021,DING2022} on that day
\begin{equation}\label{eq:carbon_cost}
  C_t = p_t^{CO_2} I_t.
\end{equation}
Daily carbon spot prices are collected from \cite{EUETS_spot_data}.

\begin{figure}[H]
    \centering
    \includegraphics[width=1\textwidth]{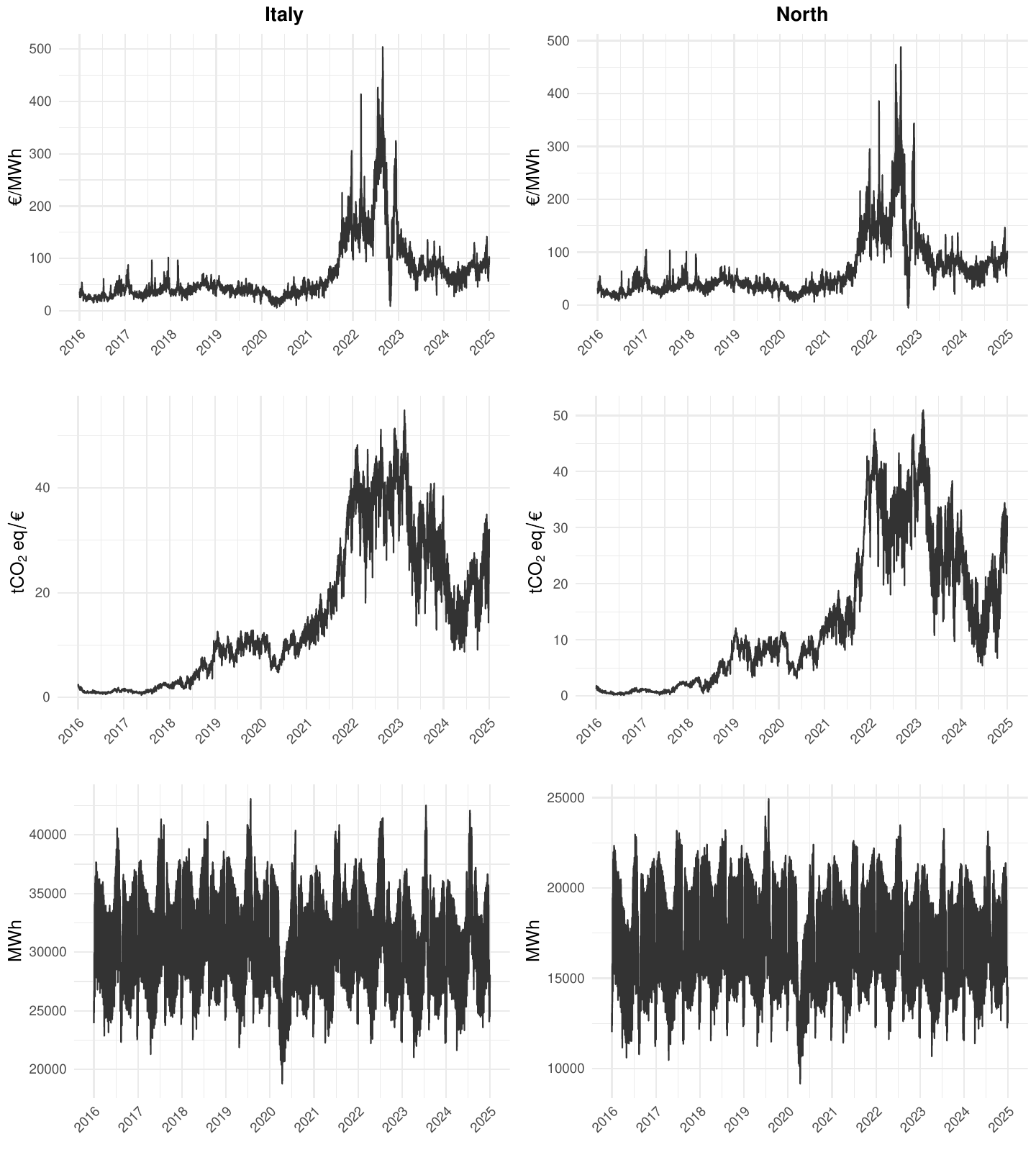}
    \caption{Daily spreads between electricity prices and fuel costs (top panels), carbon costs (middle panels), and electricity demand (bottom panels) for Italy and the North zone, from 1 January 2016 to 31 December 2024.}
    \label{fig:series.italy}
\end{figure}

Figure \ref{fig:series.italy} shows the three main variables used to estimate the CPTR at a daily frequency for Italy and the North zone from 1 January 2016 to 31 December 2024. Panel A displays the spread $s_t$ between the volume-weighted wholesale electricity spot price and the fuel cost of power generation, Panel B shows the carbon cost $C_t$, and Panel C presents electricity demand $D_t$. The spread remains relatively stable from 2016 to 2019, falls sharply during the 2020 pandemic, rises strongly in 2021-2022 amid the energy crisis, and gradually returns to lower levels in 2023-2025. The carbon cost is low and flat in 2016-2017, increases steadily from 2018, reaches historically high levels in 2021-2022, and stabilizes thereafter at an elevated level. Electricity demand exhibits strong daily, weekly, and annual seasonal patterns, with a dip in 2020 followed by a partial recovery. Figures \ref{fig:series.cnorth.csouth} and \ref{fig:series.sicily.sardinia} in Appendix \ref{sec.appendix.fig} display the three variables the other market zones.

\begin{table}[H]
\centering
\begin{tabular}{lcccccccc}
\hline
Market               & Obs                  & Mean                 & Median               & SD                   & Max                  & Min                  & Skewness             & Kurtosis             \\ \hline
\multicolumn{9}{l}{Entire sample period (1 January, 2016 - 31 December, 2024)} \\
Italy                & 3288 & 71.57 & 47.77 & 62.17 & 504.09 & 5.69 & 2.64 & 8.62 \\
North & 3288 & 67.61 & 46.15 & 58.86 & 488.24 & -5.19 & 2.66 & 9.21 \\ 
Centre-North    & 3289 & 67.00 & 44.79 & 57.67 & 482.75 & -10.84 & 2.66 & 9.38 \\
Centre-South    & 3288 & 75.35 & 49.78 & 65.45 & 527.72 & 6.12 & 2.58 & 8.12 \\
Sicily   & 3288 & 77.67 & 57.64 & 61.36 & 535.00 & -16.89 & 2.64 & 9.08 \\ 
Sardinia   & 3278 & 89.21 & 55.73 & 83.48 & 660.92 & 4.55 & 2.59 & 7.72 \\ 
\multicolumn{9}{l}{EU ETS Phase~3 (1 January, 2016 - 31 December, 2020)} \\
Italy      & 1827 & 37.61 & 36.97 & 11.51 & 102.16 & 5.69 & 0.71 & 1.88 \\ 
North     & 1827 & 35.73 & 34.64 & 12.85 & 105.16 & 4.95 & 0.93 & 2.38 \\ 
Centre-North  & 1828 & 35.17 & 34.23 & 11.42 & 101.23 & 5.10 & 0.92 & 2.52 \\ 
Centre-South  & 1827 & 39.66 & 39.09 & 10.99 & 107.38 & 6.12 & 0.68 & 2.66 \\ 
Sicily    & 1827 & 45.86 & 44.38 & 15.52 & 116.32 & -5.88 & 0.43 & 0.42 \\ 
Sardinia   & 1817 & 44.82 & 44.31 & 12.73 & 110.48 & 4.55 & 0.40 & 1.51 \\ 
\multicolumn{9}{l}{EU ETS Phase~4 (1 January, 2021 - 31 December, 2024)}  \\
Italy   & 1461 & 114.03 & 88.66 & 72.72 & 504.09 & 8.74 & 1.92 & 4.21 \\ 
North    & 1461 & 107.48 & 84.87 & 68.78 & 488.24 & -5.19 & 2.00 & 4.87 \\ 
Centre-North    & 1461 & 106.83 & 87.11 & 66.87 & 482.75 & -10.84 & 2.03 & 5.23 \\ 
Centre-South     & 1461 & 119.99 & 91.51 & 76.85 & 527.72 & 17.18 & 1.81 & 3.74 \\ 
Sicily & 1461 & 117.44 & 96.04 & 72.97 & 535.00 & -16.89 & 1.91 & 4.44 \\ 
Sardinia        & 1461 & 144.42 & 108.41 & 99.68 & 660.92 & 13.53 & 1.72 & 3.02 \\  \hline
\end{tabular}
\caption{Descriptive statistics of the spread for the sample period (1 January, 2016 - 31 December, 2024) as well as for the EU ETS Phase~3 (1 January, 2016 - 31 December, 2020), and the EU ETS Phase~4 (1 January, 2021 - 31 December, 2024).}
\label{tab.desc.spread}
\end{table}

Table \ref{tab.desc.spread} reports descriptive statistics of the daily spread for the full sample period, as well as for EU ETS Phases 3 and 4. Across all markets, daily spreads exhibit high standard deviations and kurtosis. The spread distributions are right-skewed, with the mean generally exceeding the median across all markets and periods. Over the full sample period, Italy records an average daily spread of 71.57 \texteuro{}/MWh. Among the zonal markets, Sardinia shows the highest spreads, with an average of 89.21 \texteuro{}/MWh, followed by Sicily (77.67 \texteuro{}/MWh) and Centre-South (75.35 \texteuro{}/MWh). Lower average spreads are observed in North (67.61 \texteuro{}/MWh) and Centre-North (67.00 \texteuro{}/MWh). This ranking reflects differences in generation mixes and market conditions across zones.

Table \ref{tab.desc.spread}, Figure \ref{fig:series.italy}, and Figures \ref{fig:series.cnorth.csouth}--\ref{fig:series.sicily.sardinia} highlight the substantial temporal variation in spreads. During EU ETS Phase~4, the daily average Italian spread increases to 114.03 \texteuro{}/MWh, with a maximum value of 504.09 \texteuro{}/MWh. Spreads are significantly lower in Phase~3 compared with Phase~4. In Italy, the average spread rises from 37.61 \texteuro{}/MWh in Phase~3 to 114.03 \texteuro{}/MWh in Phase~4, corresponding to an increase of over 200\%. This sharp increase is consistent across all zones and may reflect tighter EU ETS allowance constraints, as well as broader global factors, including the surge in energy prices following the Russia-Ukraine conflict \citep{SUN2024}.

Table~\ref{tab.mean_markets_transposed} reports the average carbon intensity across Italian market zones over the entire sample period and by EU ETS phases. Carbon intensity increases from Phase~3 to Phase~4 in most zones, particularly in the North (from 0.28 to 0.39) and Sardinia (from 0.70 to 0.94), indicating a higher reliance on carbon-intensive generation. In contrast, the Centre--South and Sicily show a reduction (from 0.47 to 0.38 and from 0.64 to 0.52, respectively), reflecting the growing contribution of renewable energy and the phase-out of coal \citep{MASE2024PNIEC}. Overall, the results highlight significant zonal heterogeneity in the evolution of carbon intensity across Italy.\\
Part of the increase observed in Phase~4 can be linked to emergency policy measures adopted during the gas crisis. Article~5-bis of Decree-Law~14/2022 introduced a temporary programme aimed at reducing natural gas consumption by maximizing the utilization of coal- and oil-fired thermoelectric plants \citep{arera270_2024}. Within this framework, Terna defined and updated a weekly dispatch schedule for plants with capacity above 300~MW, ensuring their maximum utilization subject to system security constraints. The programme was implemented between September~2022 and March~2023, thus falling within Phase~4.

\begin{table}[H]
\centering
\begin{tabular}{lcccccc}
\hline
 & Italy & North & Centre-North & Centre-South & Sicily & Sardinia \\ \hline
Entire sample period & 0.38 & 0.33 & 0.26 & 0.43 & 0.59 & 0.80 \\
EU ETS Phase~3       & 0.34 & 0.28 & 0.25 & 0.47 & 0.64 & 0.70 \\
EU ETS Phase~4       & 0.43 & 0.39 & 0.27 & 0.38 & 0.52 & 0.94 \\
\hline
\end{tabular}
\caption{Average carbon intensity for the sample period (1 January, 2016 - 31 December, 2024) as well as for the EU ETS Phase~3 (1 January, 2016 - 31 December, 2020), and the EU ETS Phase~4 (1 January, 2021 - 31 December, 2024).}
\label{tab.mean_markets_transposed}
\end{table}

\section{Methodology}\label{sec.methodology}
This section describes the methodology used to estimate the CPTR and some preliminary issues. 

\subsection{Stationary and autocorrelation issues}
\begin{figure}[H]
    \centering
    \includegraphics[width=0.95\linewidth]{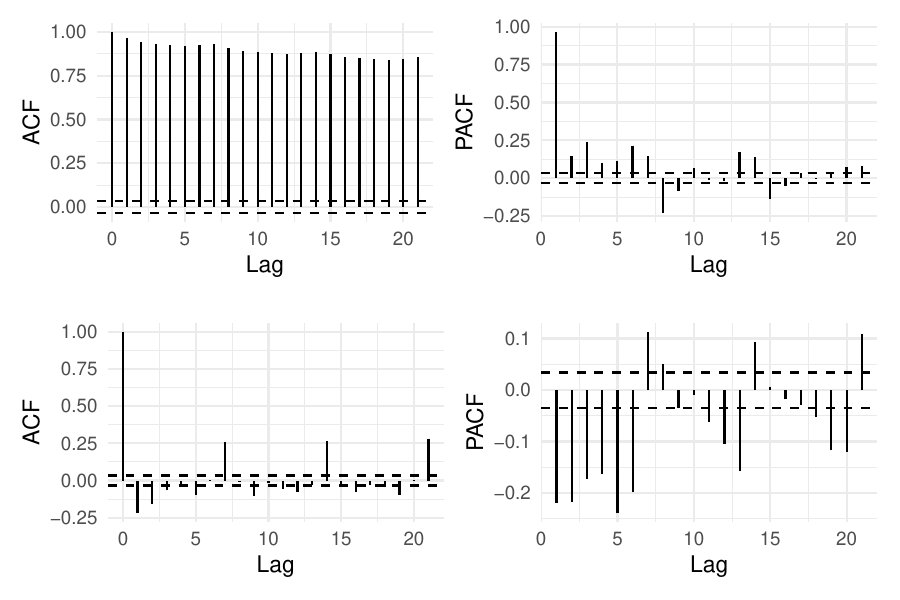}
    \caption{Autocorrelation (ACF) and partial autocorrelation (PACF) functions of the spread for Italy. The top panels show the ACF and PACF of the spread in levels, while the bottom panels show those of the log-differenced spread.}
    \label{fig:fig.acf.italy}
\end{figure}

Figure \ref{fig:series.italy} and Figures \ref{fig:series.cnorth.csouth}--\ref{fig:series.sicily.sardinia} suggest that the series in levels exhibit apparent non-stationarity. To formally investigate their integration properties, the Augmented Dickey-Fuller (ADF) test \citep{Dickey1979} and the Kwiatkowski-Phillips-Schmidt-Shin (KPSS) test \citep{KWIATKOWSKI1992} are applied. The ADF test has the null hypothesis of non-stationarity, while the KPSS test assumes that the series are stationary under the null. Table \ref{tab:unirootitaly} reports the results for Italy. The ADF test indicates that the spread and carbon costs are non-stationary, whereas the KPSS test suggests that the spread, carbon cost, and average electricity demand are non-stationary.
\begin{table}[H]
\centering
\begin{tabular}{lcccccc}
\hline
 & $s_t$ & $C_t$ & $D_t$ & $\tilde s_t$ & $\tilde C_t$ & $\log(D_t)$ \\ 
\hline
ADF  & -3.08 & -2.34 & -7.48* & -20.86* & -18.88* & -7.32* \\ 
KPSS & 13.18* & 5.63* & 36.59* & 0.01 & 0.29 & 0.94 \\ 
\hline
\end{tabular}
\caption{Unit root tests for Italy. Significance level: p-value $\leq 0.05$.}
\label{tab:unirootitaly}
\end{table}
The top panels of Figure \ref{fig:fig.acf.italy} shows the ACF and PACF of the spread in levels for Italy. The ACF decays slowly, indicating strong persistence and non-stationarity, while the PACF displays a dominant first-lag spike followed by smaller significant coefficients, consistent with an autoregressive process.\\ Since the series in levels are non-stationary, the empirical analysis is conducted using first differences of logarithms \citep{Wooldridge2008}. The transformed variables are defined as follows: the first difference of log prices, $\tilde{s}_t = \Delta \big(\log p_t^e - \log p_t^f\big)$; the first difference of log carbon costs, $\tilde{C}t = \Delta \log C_t = \log C_t - \log C{t-1}$; and log demand, $\log D_t$. The transformed variables are stationary according to both unit root tests in Table \ref{tab:unirootitaly}. The log-difference specification implies an elasticity-based interpretation, rather than the conventional level-based CPTR interpretation \citep{NAZIFI2016,Caporin2021}. The estimated coefficients measure the percentage change in the electricity-fuel price spread associated with a one percent change in carbon costs. Hence, they should be interpreted as carbon cost transmission elasticities rather than strict level-based CPTR. For ease of exposition, the term CPTR is used to denote this elasticity-based measure of carbon cost transmission.

The bottom panels of Figure \ref{fig:fig.acf.italy} shows the ACF and PACF of the log-differenced spread for Italy, which appears stationary. The ACF decays rapidly, although notable spikes remain at lags 7, 14, and 21, indicating weekly seasonal patterns. The PACF displays significant coefficients mainly within the first seven lags, suggesting short-run dynamics. The same conclusions hold for the other market zones.
\subsection{Causality and cointegration issues}
To investigate whether carbon costs contain predictive information for the dynamics of the electricity--fuel spread, Appendix~\ref{sec.appendix.causality} reports standard Granger tests, Toda--Yamamoto tests, and VECM-based short-run, long-run, and joint causality tests \citep{Granger1969,TodaYamamoto1995,Johansen1991,EngleGranger1987}. The results indicate that carbon costs contain predictive information for the electricity-fuel price spread in the national market and in most bidding zones, thereby supporting the existence of a causal relationship. Standard Granger and Toda--Yamamoto tests reject the null in all zones. The VECM-based results further indicate short-run causality, long-run adjustment, and joint causality in all zones for which a cointegrating relation is detected, with the exception that VECM statistics are not reported for Sardinia because no positive cointegration rank is selected.

Non-stationarity and autocorrelation violate key assumptions underlying OLS estimation, leading to biased and unreliable estimates of the CPTR. In the existing literature, these issues have been overlooked, with the exception of \cite{NAZIFI2016}, who explicitly investigates stationarity of the variables, and \cite{Caporin2021}, who employ differenced variables and VECM models.\\
In this work, the proposed model addresses both concerns by ensuring stationarity through (log-)differencing and by incorporating an autoregressive structure to capture persistence, thereby mitigating residual autocorrelation.\\
Following \cite{Wooldridge2008}, transforming variables into (log-)differences removes stochastic trends but does not imply a zero mean; rather, the transformed series exhibits a constant mean growth rate. Importantly, this transformation does not affect the interpretation of the CPTR coefficients. Under the log-difference specification, the coefficients $(\beta_1, \beta_2)$ capture the response of changes in electricity prices to changes in carbon costs, that is, pass-through elasticities.\\
Finally, potential heteroskedasticity in the residuals is addressed by employing heteroskedasticity and autocorrelation consistent standard errors, computed using the Newey-West estimator.

\subsection{CPTR estimation}
To estimate the impact of carbon prices on Italian spot electricity prices, an OLS regression model is employed, following \cite{CHEN2008,SIJM2006,SIJM2008,NAZIFI2016,NAZIFI2021,DING2022}. In addition to the standard specification, lagged dependent variables are included to capture persistence and stochastic seasonal dynamics in electricity prices:
\begin{equation}\label{eq:cptr.model}
\tilde s_t
= \beta_0
+ \sum_{\ell} \phi_{\ell} \tilde s_{t-\ell}
+ \beta_1 \tilde C_t
+ \beta_2 \tilde C_t I_t
+ \beta_3 \log(D_t)
+ \epsilon_t,
\end{equation}
where $\ell \in \{1,2,3,4,5,7,14,21\}$. The intercept $\beta_0$ reflects the fixed component of the price spread, such as generator operating costs \citep{NAZIFI2021}.\\
Short lags ($\ell = 1,\ldots,5$) capture short-run persistence, while $\ell = 7,14,21$ account for stochastic seasonality.\\
The term $I_t$ is a dummy variable equals 1 during Phase~4 and 0 otherwise. The coefficient $\beta_1$ measures the CPTR during Phase~3, while $\beta_2$ captures the change in this effect in Phase~4, implying a Phase~4 CPTR of $(\beta_1 + \beta_2)$. A CPTR equal to 1 indicates full pass-through of carbon costs to electricity prices; values below (above) 1 imply incomplete pass-through (over-shifting), with generators absorbing part of the cost (or consumers bearing more than the direct cost) \citep{NAZIFI2021}. The coefficient $\beta_3$ captures the effect of demand events on electricity prices. The error term $\epsilon_t$ captures the effect of unobserved factors.

Since spot electricity prices typically exhibit a non-linear relationship with demand, robustness checks are conducted using both polynomial specifications and a smooth demand function. First, quadratic and cubic terms in log demand, $\log(D_t)^2$ and $\log(D_t)^3$, are added to Eq.~(\ref{eq:cptr.model}). Second, a generalized additive model (GAM) is specified, allowing the fuel spread to depend on a smooth function of demand \citep{DING2022,NAZIFI2021}:
\begin{equation}\label{eq:GAM}
\tilde s_t
= \beta_0
+ \sum_{\ell} \phi_{\ell} \tilde s_{t-\ell}
+ \beta_1 \tilde C_t
+ \beta_2 \tilde C_t \, \mathbb{I}_t
+ f\!\left(\log(D_t)\right)
+ \epsilon_t,
\end{equation}
where $f(\log(D_t))$ is approximated by a set of basis functions $\{\psi_k(\cdot)\}_{k=1}^K$ such that
\[
f(\log(D_t)) = \sum_{k=1}^K \Psi_k \, \psi_k\!\left(\log(D_t)\right).
\]
A cubic spline basis is employed, and the model is estimated with the \textsf{R} package \texttt{mgcv}.

A quantile regression (QR) model \citep{Koenker1978} is employed to examine how the CPTR varies across different points of the conditional distribution of the  spread:
\begin{equation}\label{eq:QR}
Q_\tau(\tilde s_t \mid \tilde C_t, \mathbb{I}_t, \log(D_t)) 
= \beta_0(\tau)
+ \sum_{\ell} \phi_{\ell}(\tau) \tilde s_{t-\ell}
+ \beta_1(\tau) \tilde C_t
+ \beta_2(\tau) \tilde C_t \, \mathbb{I}_t
+ \beta_3(\tau) \log(D_t),
\end{equation}
where $Q_\tau(\tilde s_t \mid \tilde C_t, \mathbb{I}_t, \log(D_t))$ is the $\tau$-th $0<\tau<1$ quantile of the spread, conditional on $\tilde C_t$, $\mathbb{I}_t$, and $\log(D_t)$. Lower (higher) values of $\tau$ correspond to the lower (upper) tail of the distribution, allowing the CPTR to differ across the distribution of the spread.

\section{Empirical results}\label{sec.results}
\subsection{The baseline model}

\begin{figure}[H]
    \centering
    \includegraphics[width=0.98\linewidth]{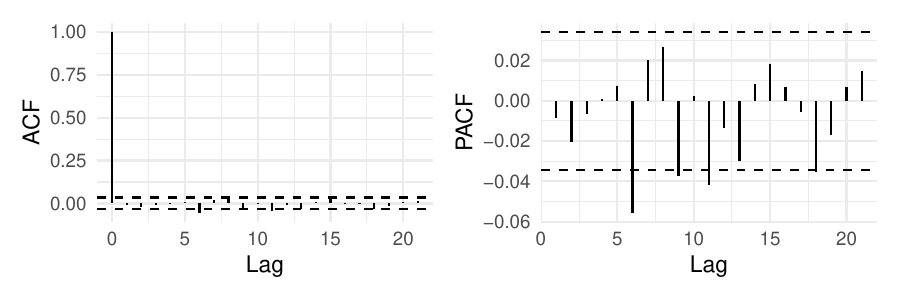}
    \caption{Autocorrelation (ACF) and partial autocorrelation (PACF) functions of the residuals for the estimated regression model for Italy.}
    \label{fig:fig.acf.resitaly}
\end{figure}

Tables \ref{tab:cptr_estimates}-\ref{tab:phase_impacts} report the carbon pass-through rates and results for the applied regression model in Eq. (\ref{eq:cptr.model}). The first five autoregressive coefficients $\phi_1$--$\phi_5$ are negative and highly significant across all zones, indicating rapid short-term adjustment. In contrast, the coefficients $\phi_7$, $\phi_{14}$, and $\phi_{21}$ are positive and significant, confirming the presence of stochastic seasonality. The inclusion of adjacent lagged dependent variables does not generate problematic multicollinearity. Table~\ref{tab:vif_baseline} in Appendix \ref{sec.appendix.tab} reports the variance inflation factors (VIF) for the autoregressive terms. All VIFs are low ranging from 1.05 to 1.53 across market zones.\\ Figure \ref{fig:fig.acf.resitaly} shows the ACF and PACF of the residuals for the estimated regression model for Italy. The autocorrelations are close to zero and lie within the confidence bands at all lags, indicating no significant serial correlation.\\
Electricity demand ($\beta_3$) has a positive and significant effect in most markets, although it is not statistically significant in the Centre--South.

At the national level, the estimated carbon cost pass-through rate is positive and statistically significant in Phase~3, with $\hat{\beta}_1 = 0.32$. The interaction term $\hat{\beta}_2$ is small and not statistically significant. The aggregate CPTR remains stable across the two phases, with no evidence of a statistically significant change for the transition from Phase~3 to Phase~4.\\
At the zonal level, substantial heterogeneity emerges. The estimated CPTRs are positive and statistically significant during Phase~3 across all zones. The additional effect in Phase~4 is positive and statistically significant for the North, Centre-North, and Sardinia, whereas it is negative for the Centre-South and Sicily. The magnitude of these effects varies across zones. In Phase~3, Sicily records the highest CPTR (41\%), followed by the Centre-South (20\%), the Centre-North (18\%), and the North (17\%).
In Phase~4, Sardinia leads with the highest CPTR (41\%), followed by the Centre-North and the North (both 24\%), Sicily (20\%), and the Centre-South (5\%).

The economic theory states that marginal production costs increase proportionally with carbon costs. These costs should be fully passed through to electricity prices under perfect competition, inelastic demand, and perfectly elastic supply \citep{SIJM2006,SIJM2008,SIJM2012}. In this framework, any increase in marginal costs induced by carbon pricing is reflected in market prices. The estimated CPTR suggest incomplete pass-through \citep{NAZIFI2021}. This finding may indicate that part of the carbon costs is absorbed by electricity generators rather than being fully transferred to consumers. The increases in carbon intensity in the North, Centre--North, and Sardinia between Phase~3 and Phase~4 are associated with an increase in CPTR over the same period. By contrast, in the Centre--South and Sicily, a decrease in carbon intensity coincides with a reduction in CPTR.\\
Several factors may help to explain the heterogeneity observed in the CPTR estimates across the bidding market zones.\\

\begin{table}[H]
\begin{center}
\begin{tabular}{lcccccc}
\hline
\multicolumn{1}{l}{\rule{0pt}{12pt}}
 & Italy & North & Centre--North & Centre--South & Sicily & Sardinia \\[2pt]
\hline\rule{0pt}{12pt}$\beta_0$ & -1.86** & -2.03** & -0.93** & 0.02 & -1.01** & -1.17** \\
$\phi_1$ & -0.38** & -0.37** & -0.31** & -0.39** & -0.34** & -0.35** \\
$\phi_2$ & -0.28** & -0.28** & -0.25** & -0.33** & -0.22** & -0.29** \\
$\phi_3$ & -0.22** & -0.23** & -0.19** & -0.25** & -0.18** & -0.22** \\
$\phi_4$ & -0.16** & -0.16** & -0.14** & -0.21** & -0.13** & -0.16** \\
$\phi_5$ & -0.12** & -0.14** & -0.14** & -0.18** & -0.10** & -0.11** \\
$\phi_7$ & 0.04* & 0.08** & 0.10** & 0.05** & 0.05** & 0.08** \\
$\phi_{14}$ & 0.05** & 0.07** & 0.09** & 0.08** & 0.09** & 0.10** \\
$\phi_{21}$ & 0.09** & 0.11** & 0.12** & 0.08** & 0.10** & 0.13** \\
$\beta_1$ & 0.32** & 0.17** & 0.18** & 0.20** & 0.41** & 0.12** \\
$\beta_2$ & -0.03 & 0.07* & 0.06** & -0.15** & -0.21** & 0.29** \\
$\beta_3$ & 0.18** & 0.21** & 0.12** & -0.003 & 0.13** & 0.17** \\
\hline
$R_{Adj}^2$ & 39.48 & 46.59 & 46.15 & 25.41 & 34.85 & 26.32 \\[2pt]
\hline
\end{tabular}
\end{center}
\caption{Carbon pass-through rates and results for the applied regression model.
Significance levels: ** p-value$\leq$0.01, * p-value$\leq$0.05.}
\label{tab:cptr_estimates}
\end{table}

\begin{table}[H]
\begin{center}
\resizebox{\linewidth}{!}{
\begin{tabular}{lcccccc}
\hline
\multicolumn{1}{l}{\rule{0pt}{12pt}}
& Italy & North & Centre--North & Centre--South & Sicily & Sardinia \\[2pt]
\hline\rule{0pt}{12pt}EU ETS Phase~3 & 0.32 & 0.17 & 0.18 & 0.20 & 0.41 & 0.12 \\
EU ETS Phase~4 & 0.29 & 0.24 & 0.24 & 0.05 & 0.20 & 0.41 \\
\% Variation & -9.38\% & 41.18\% & 33.33\% & -75.00\% & -51.22\% & 241.67\% \\
\hline
\end{tabular}
}
\end{center}
\caption{Estimated CPTR in Phase~3 and Phase~4 with the applied regression model.}
\label{tab:phase_impacts}
\end{table}

The North zone exhibits a moderate increase in CPTR ($41.18\%$) in Phase~4. This increase occurs alongside a sharp reduction in hydroelectric generation between 2021 and 2023. In 2022, hydroelectric generation declined by 38.57\% compared to the average value of the years 2016--2021. Over the same period, thermoelectric generation from coal-fired plants increased, following government measures introduced to mitigate the surge in gas prices \citep{MASE2024PNIEC}. Although installed renewable capacity and renewable generation expanded between Phase~3 and Phase~4 (Tables~\ref{tab.capacity_split_var}--\ref{tab.production_split_var}), this increase does not appear to have fully offset the decline in hydroelectric output.\\

\begin{table}[H]
\centering
\begin{tabular}{lcccccc}
\hline
 & \multicolumn{2}{c}{EU ETS Phase~3} & \multicolumn{2}{c}{EU ETS Phase~4} & \multicolumn{2}{c}{\% Variation} \\
Zone & Wind & Solar & Wind & Solar & Wind (\%) & Solar (\%) \\ \hline
North  & 146.13 & 9649.72 & 214.11 & 17656.05 & 46.52 & 82.98 \\
Centre-North & 162.74 & 1984.73 & 164.12 & 2982.39 & 0.85 & 50.27 \\
Centre-South & 2085.74 & 3547.45 & 2533.05 & 6667.17 & 21.45 & 87.96 \\
Sicily       & 2013.04 & 1541.66 & 2489.86 & 2673.94 & 23.69 & 73.45 \\
Sardinia     & 1087.52 & 973.52 & 1193.52 & 1722.09 & 9.75 & 76.89 \\
\hline
\end{tabular}
\caption{Installed capacity of renewable sources (MWh), split between solar and wind, with percentage variation between EU ETS Phase~3 and Phase~4.}
\label{tab.capacity_split_var}
\end{table}

\begin{table}[H]
\centering
\begin{tabular}{lcccccc}
\hline
 & \multicolumn{2}{c}{EU ETS Phase~3} & \multicolumn{2}{c}{EU ETS Phase~4} & \multicolumn{2}{c}{\% Variation} \\
Zone & Wind & Solar & Wind & Solar & Wind (\%) & Solar (\%) \\ \hline
North        & 229.32 & 9502.69 & 378.86 & 12584.99 & 65.22 & 32.44 \\
Centre-North & 260.31 & 2213.40 & 303.25 & 2603.93 & 16.50 & 17.64 \\
Centre-South & 2762.59 & 4023.78 & 4293.95 & 5128.04 & 55.42 & 27.45 \\
Sicily       & 3036.88 & 1845.93 & 3472.83 & 2282.73 & 14.35 & 23.66 \\
Sardinia     & 1780.26 & 997.64  & 1811.72 & 1470.99 & 1.77  & 47.45 \\
\hline
\end{tabular}
\caption{Average renewable electricity production (GWh), split between wind and solar, with percentage variation between EU ETS Phase~3 and Phase~4.}
\label{tab.production_split_var}
\end{table}

The Centre--North also shows a moderate increase in CPTR, rising by $33.33\%$. In this zone, hydroelectric generation declined by about $45\%$, while electricity prices were influenced by higher costs for gas-fired generation.\\
In Sardinia, the CPTR increased by $241.67\%$ from Phase~3 to Phase~4. Despite some growth in renewable capacity and generation between Phase~3 and Phase~4, the system remains heavily dependent on thermal generation, with coal and bioenergy accounting for a large share of electricity production. In addition, the limited interconnection capacity with the mainland, may constrain imports of lower-cost electricity, particularly during peak periods \citep{TernaSAPEI,SAPIO2020}.\\
The Centre--South and Sicily show a marked reduction in CPTR, of $75\%$ and $51.22\%$, respectively. These reductions coincide with substantial growth in renewable capacity and generation between Phase~3 and Phase~4. In the Centre--South, solar capacity increased by 87.96\%, while wind capacity grew by 21.45\%. Similarly, in Sicily, solar capacity expanded by about 73.45\% and wind capacity by 23.69\% \citep{Terna_Generazione_2025}. This expansion of low-marginal-cost renewable energy may have reduced the role of thermal generation in price formation. In the Centre--South, solar generation increased by 23.66\%, while wind generation grew by 55.42\%. Similarly, in Sicily, solar generation expanded by about 23.66\% and wind generation by 14.25\% \citep{Terna_Generazione_2025}. In addition, the Centre--South experienced a sharp decline in coal-based generation, which fell from over 12 TWh in 2016 to less than 0.4 TWh in 2024 \citep{Terna_Generazione_2025}.\\ 
Overall, these factors provide a plausible context for the observed reduction in CPTRs in these zones, although they should not be interpreted as a direct causal explanation.

A specification based on average daily prices is also considered as an alternative to volume-weighted electricity prices. Additional robustness checks include models with quadratic and cubic terms in log demand, a generalized additive model (GAM) in which the fuel spread depends on a smooth function of demand, a specification controlling for the coal-maximization programme implemented during the energy crisis, and a model including calendar fixed effects.

The results based on average daily electricity prices (Table~\ref{tab:cptr_estimates_dailyavg} in Appendix~\ref{sec.appendix.tab} are very similar to those of the baseline model, indicating that the estimated carbon pass-through rates are robust to the choice of price measure. Allowing for nonlinear demand effects through a quadratic term (Table~\ref{tab:cptr_estimates_quad} in Appendix~\ref{sec.appendix.tab}) does not affect the results. The estimated carbon pass-through rates remain close to those in the baseline specification, with $\beta_1$ and $\beta_2$ largely unchanged across market zones. The quadratic term suggests only negligible nonlinearities except for the Centre--South, and the adjusted $R^2$ shows no meaningful improvement. A similar pattern emerges from the GAM specification (Table~\ref{tab:cptr_estimates_gam} in Appendix~\ref{sec.appendix.tab}). Although the smoothing term is statistically significant and points to mild nonlinearities in most market zones, the estimated pass-through rates and the overall fit remain very close to those of the linear model.

The additional checks reported in Tables~\ref{tab:coal_max_dummy} and~\ref{tab:week_month_fixed_effects} in Appendix~\ref{sec.appendix.tab}) further support the robustness of the baseline estimates. Controlling for the temporary coal-maximization programme implemented between September 2022 and March 2023 leaves the estimated coefficients almost unchanged relative to the baseline model. The carbon cost pass-through coefficient in Phase 3, $\beta_1$, remains positive and statistically significant in all zones, with values identical or extremely close to those reported in the baseline specification. Similarly, the Phase 4 differential effect, $\beta_2$, preserves both its sign and statistical significance across market zones. This suggests that the heterogeneous change in pass-through between Phase 3 and Phase 4 is not driven by the emergency policy aimed at increasing coal- and oil-fired generation. The coefficient associated with the coal-maximization dummy, $\beta_4$, is small in magnitude and statistically insignificant in most zones, becoming weakly significant only in Centre--North.

The specification including day-of-week and month fixed effects provides further evidence that the baseline estimates are not driven by recurring calendar seasonality. After controlling for systematic weekly and monthly patterns, the estimates of $\beta_1$ remain positive and highly significant across all zones. The magnitude of the pass-through coefficient is only slightly lower than in the baseline model for Italy, North, Centre--North, and Sardinia, while it remains essentially unchanged in Sicily and slightly increases in Centre--South. The Phase 4 interaction term, $\beta_2$, is also broadly consistent with the baseline results, confirming an increase in the pass-through rate in Sardinia and Centre--North and a decrease in Centre--South and Sicily. The only notable difference concerns the North, where the Phase 4 differential effect becomes smaller and loses statistical significance once calendar fixed effects are included.

The calendar specification also helps clarify the role of the autoregressive terms. In the baseline model, the coefficients at lags 7, 14, and 21 days are positive and statistically significant in most zones, suggesting the presence of weekly seasonal dependence in the fuel spread. Once day-of-week and month fixed effects are included, these seasonal lag coefficients become smaller and lose statistical significance in most cases. This is expected because the baseline autoregressive structure captures part of the seasonal component stochastically through lagged values of the dependent variable, while the calendar fixed effects model these recurring patterns directly through deterministic dummy variables. Therefore, the lower significance of some $\phi$ coefficients in the calendar specification does not weaken the baseline model; instead, it confirms that part of the serial dependence at weekly horizons reflects regular calendar seasonality.

Overall, these results confirm the robustness of the baseline findings and support the adequacy of the parsimonious linear specification.

\subsection{The quantile regression model}
\begin{figure}[H]
    \centering
    \includegraphics[width=1\linewidth]{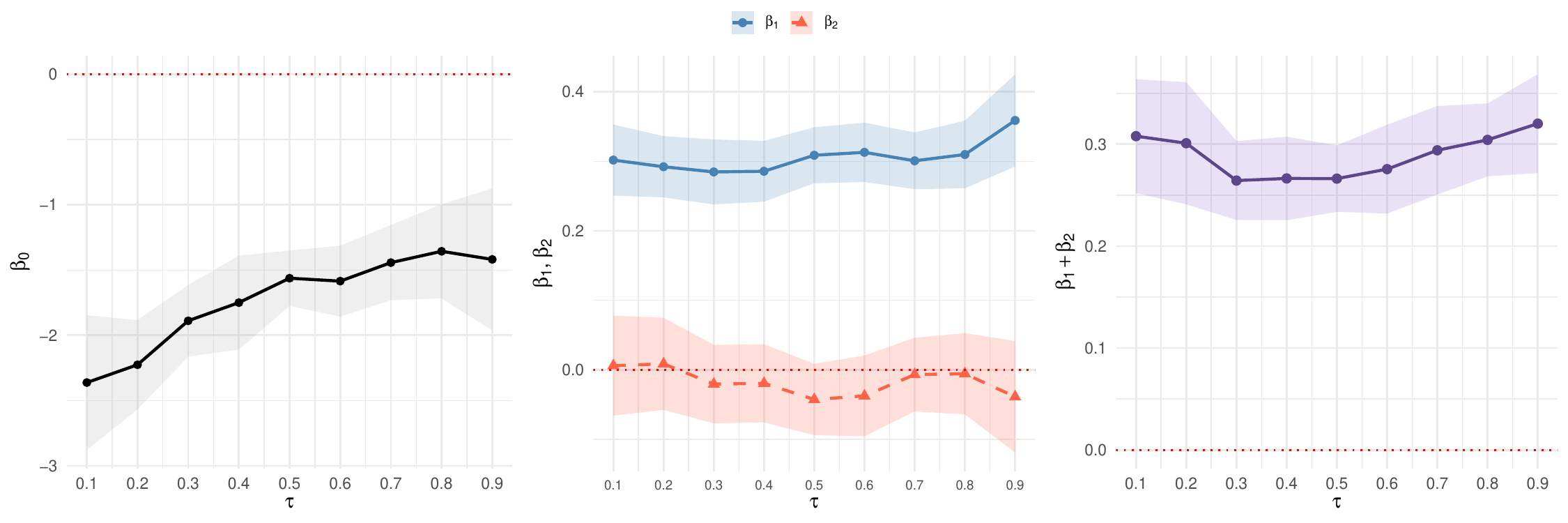}
    \caption{Quantile regression estimates for Italy. The estimated coefficients $\beta_0$ (left panel), $\beta_1$ and $\beta_2$ (middle panel), and their sum $\beta_1 + \beta_2$ (right panel) are plotted across quantiles $\tau$. Shaded areas indicate $90\%$ confidence intervals.}
    \label{fig.quant.italy}
\end{figure}

To further investigate CPTR behaviour, its variation across the price spread distribution is analysed, i.e., whether it changes with the price level. To do that a quantile regression model is applied.\\
First the estimates of $\beta_0(\tau)$ point out a positive, significant, correlation between fixed costs, represented by $\beta_0(\tau)$, and the price spread quantile. This suggest that, not surprisingly, an increase in fixed costs translates in a, roughly linear, increase in prices.\\
Table \ref{tab:qr_betas_all_zones} reports the quantile regression estimates of $\beta_0(\tau)$, $\beta_1(\tau)$, and $\beta_2(\tau)$ for $\tau = 0.10, \ldots, 0.90$. Estimates for the remaining coefficients in Eq.~(\ref{eq:QR}) are available from the author upon request.\\
As concerns CPTR  the results show that the CPTR associated with EU ETS Phase~3, described by $\beta_1(\tau)$, is significantly positive across all quantiles and zones. Coherently with baseline model results, $\beta_1(\tau)$ is always largely less than one, meaning that only a fraction of the EU ETS costs are passed to prices, independently of their level.  
While estimates at the median ($\tau = 0.5$) closely match standard regression results, some slightly decreasing patterns, with respect to the price levels, emerge for Centre-South, Sardinia and Sicily. However, when considering the $90\%$ confidence intervals, no clear systematic differences emerge across the distributions of all zones, indicating that Phase~3 pass-through basically does not depend on the spread level.\\
Turning to Phase~4, the additional component $\beta_2(\tau)$ at the median ($\tau = 0.5$) is also consistent with standard regression estimates. Figure \ref{fig.quant.italy}  shows the quantile regression estimates of $\beta_0$, $\beta_1$ and $\beta_2$ across the conditional distribution. It is apparent that, at the national level, $\beta_2$ is small and not statistically significant, suggesting no systematic change in the overall CPTR.\\
Nonetheless, some zonal heterogeneity emerges. The Phase~4 effect is positive in the North, Centre-North and Sardinia zones, indicating an increase in the pass-trough rate, whereas in Centre-South and Sicily $\beta_2(\tau)$ is negative, suggesting a lower pass-trough rate.
In the North, Centre-South, and Sicily, $\beta_2(\tau)$ exhibits a very mild upward pattern toward higher quantiles, partly compensating the downward behaviour of $\beta_1(\tau)$. By contrast, in the Centre-North, $\beta_2(\tau)$ exhibits a slight decline across the distribution. Sardinia represents a clear exception, with $\beta_2(\tau)$ increasing at both lower and upper quantiles. This pattern suggests that, in Sardinia, spread fluctuations have a stronger impact under bullish ($\tau > 0.50$) and bearish ($\tau < 0.50$) market conditions, while their effect is more limited around the median \citep{su17073118}.\\
When considering the global Phase~4 CPTR, given by $\beta_1(\tau)+\beta_2(\tau)$, apart from Sardinia where a parabolic pattern occurs, we can observe some compensating effect in the Phase~3 upward o downward patterns so that the Phase~4 CPTR appears more flat and, hence, not depending on the price spread levels. These considerations are also supported by the $90\%$ confidence intervals of the estimators (shaded areas in the graphs), computed using their variance--covariance matrix. In particular, for (${\beta}_1 + {\beta}_2)$ they are given by $(\hat{\beta}_1 + \hat{\beta}_2) \pm z_{0.95} \cdot \text{SE}(\hat{\beta}_1 + \hat{\beta}_2)$, where $z_{0.95} = 1.645$ and 
\begin{equation}
 \text{SE}(\hat{\beta}_1 + \hat{\beta}_2)
=
\sqrt{
\text{Var}(\hat{\beta}_1)
+
\text{Var}(\hat{\beta}_2)
+
2\,\text{Cov}(\hat{\beta}_1, \hat{\beta}_2)
}.  
\end{equation}
For the other market zones the results are shown in Figures \ref{fig.quant.north}--\ref{fig.quant.sardinia} in the Appendix \ref{sec.appendix.fig}.

\begin{table}[H]
\centering
\resizebox{\linewidth}{!}{
\begin{tabular}{llccccccccc}
  \hline
Zone    & \multicolumn{10}{c}{Quantile ($\tau$)} \\ 
 &  & $0.1$ & $0.2$ & $0.3$ & $0.4$ & $0.5$ & $0.6$ & $0.7$ & $0.8$ & $0.9$ \\ 
  \hline
\multicolumn{10}{l}{Italy} \\ 
 & $\beta_0$ & -2.36** & -2.23** & -1.89** & -1.75** & -1.56** & -1.59** & -1.44** & -1.36** & -1.42** \\ 
   & $\beta_1$ & 0.30** & 0.29** & 0.28** & 0.29** & 0.31** & 0.31** & 0.30** & 0.31** & 0.36** \\ 
   & $\beta_2$ & 0.01 & 0.01 & -0.02 & -0.02 & -0.04 & -0.04 & -0.01 & -0.01 & -0.04 \\ 
\multicolumn{10}{l}{North} \\ 
 & $\beta_0$ & -3.04** & -2.51** & -2.06** & -1.93** & -1.82** & -1.77** & -1.60** & -1.49** & -1.37** \\ 
   & $\beta_1$ & 0.18** & 0.17** & 0.15** & 0.16** & 0.15** & 0.16** & 0.18** & 0.16** & 0.17** \\ 
   & $\beta_2$ & 0.06 & 0.05 & 0.06 & 0.04 & 0.05 & 0.05 & 0.07* & 0.11** & 0.11** \\ 
\multicolumn{10}{l}{Centre-North} \\ 
 & $\beta_0$ & -1.19** & -0.95** & -0.91** & -0.95** & -0.88** & -0.82** & -0.75** & -0.72** & -0.60** \\ 
   & $\beta_1$ & 0.18** & 0.18** & 0.17** & 0.17** & 0.16** & 0.17** & 0.17** & 0.18** & 0.19** \\ 
   & $\beta_2$ & 0.09** & 0.05 & 0.05* & 0.06* & 0.08** & 0.08** & 0.06* & 0.02 & 0.01 \\ 
\multicolumn{10}{l}{Centre-South} \\ 
 & $\beta_0$ & -2.79** & -1.55** & -1.08** & -0.43 & 0.14 & 0.67** & 1.26** & 2.04** & 3.03** \\ 
   & $\beta_1$ & 0.23** & 0.20** & 0.20** & 0.20** & 0.20** & 0.21** & 0.18** & 0.16** & 0.17** \\ 
   & $\beta_2$ & -0.17** & -0.15** & -0.15** & -0.14** & -0.14** & -0.15** & -0.12** & -0.10* & -0.14** \\ 
\multicolumn{10}{l}{Sicily} \\
& $\beta_0$ & -2.46** & -1.65** & -1.18** & -0.95** & -0.77** & -0.42** & -0.21 & 0.02 & 0.85** \\ 
   & $\beta_1$ & 0.45** & 0.43** & 0.39** & 0.36** & 0.37** & 0.36** & 0.37** & 0.37** & 0.36** \\ 
   & $\beta_2$ & -0.26** & -0.21** & -0.19** & -0.17** & -0.17** & -0.17** & -0.18** & -0.18** & -0.14** \\ 
\multicolumn{10}{l}{Sardinia} \\
 & $\beta_0$ & -3.03** & -1.97** & -1.53** & -1.28** & -0.78** & -0.59** & -0.21 & -0.09 & 0.40 \\ 
   & $\beta_1$ & 0.15** & 0.11** & 0.11** & 0.11** & 0.10** & 0.10** & 0.08** & 0.07 & 0.07 \\ 
   & $\beta_2$ & 0.27* & 0.15** & 0.09* & 0.11* & 0.10 & 0.14** & 0.16** & 0.20** & 0.30** \\ 
   \hline
\end{tabular}
}
\caption{Quantile regression estimates for $\beta_0$, $\beta_1$, and $\beta_2$ across zones. Significance levels: ** p-value$\leq0.01$, * p-value$\leq0.05$.} 
\label{tab:qr_betas_all_zones}
\end{table}

\section{Policy implications}
The progressive tightening of emission caps under the EU ETS reflects the objective of reducing emissions by 62\% relative to 2005 levels by 2030 \citep{EUETS2025}. In the power sector, emissions reductions are expected to occur mainly through a switch from coal-fired generation to lower-emission alternatives such as natural gas and renewables. Italy committed to a progressive phase-out of coal-fired electricity generation by 2025, with the exception of Sardinia, under the PNIEC 2019\footnote{Piano Nazionale Integrato per l'Energia e il Clima 2019 - the 2019 National Integrated Energy and Climate Plan.}. This objective is supported by the planned expansion of renewable energy sources, together with investments in flexible generation capacity, grid infrastructure, and storage systems \citep{MASE2024PNIEC}.

\begin{figure}[H]
    \centering
    \includegraphics[width=1\linewidth]{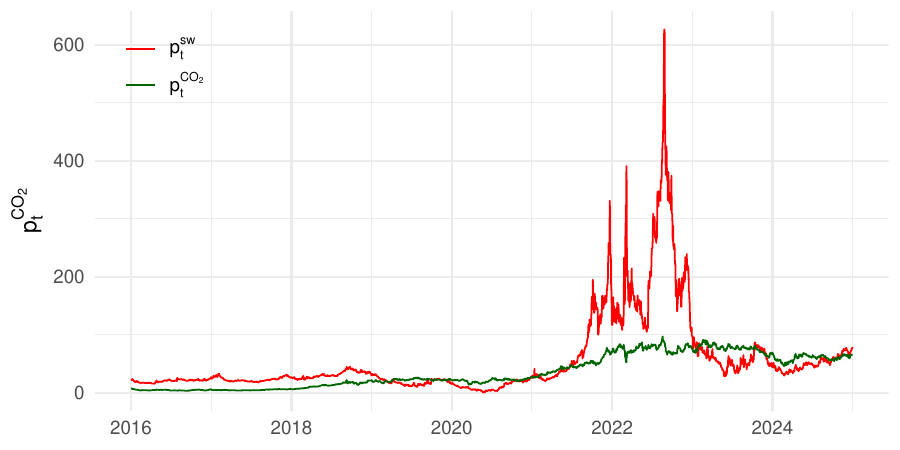}
    \caption{EU ETS carbon spot price $p_t^{CO_2}$ vs. switching price $p^{\text{sw}}_t$.}
    \label{fig:switching}
\end{figure}

To monitor market incentives to adopt less polluting technologies, the switching price is frequently used \citep{gse2025_aste_co2, LIEBENSTEINER2026}. This indicator identifies the hypothetical CO$_2$ price at which coal- and gas-fired generation become equally cost-competitive, and is defined as follows:
\begin{equation}
p^{\text{sw}}_t =
\frac{p^{g}_t - p^{c}_t}{I^{c} - I^{g}},
\end{equation}
where $p^{g}_t$ and $p^{c}_t$ denote fuel costs (in \texteuro{}/MWh), and $I^{g}$ and $I^{c}$ represent the corresponding CO$_2$ emission intensities (in tCO$_2$/MWh). When $p_t^{CO_2} < p_t^{\text{sw}}$, coal-fired generation remains more cost-competitive than gas-fired generation. Figure~\ref{fig:switching} compares the EU ETS carbon price with the switching price for Italy\footnote{The switching price for Italy is constructed using the data described in Section~\ref{sec.methodology}. Oil-fired generation is not included in the switching-price calculation because the indicator is intended to capture the main merit-order substitution channel between coal- and gas-fired generation. Although oil is considered in the construction of aggregate fuel costs and carbon intensity, its role in the Italian power system is limited to residual thermoelectric generation \cite{FRANZOSO2025}.} over the period 2016--2024. Over 2016--2019, coal remained generally more cost-competitive than gas. From early 2019 to mid-2021, this relationship reversed, with gas becoming the more competitive technology. However, between July 2021 and December 2022, the surge in gas prices during the energy crisis made gas-fired generation more expensive than coal, increasing wholesale electricity prices \citep{LIEBENSTEINER2026}. During this period, the carbon price would have needed to be approximately 2.7 times higher to restore the competitiveness of gas.\\ In Italy, the supply side was simultaneously affected by a sharp contraction in hydroelectric generation (36.3\%), driven by prolonged drought conditions. This decline was partially offset by an increase in thermoelectric generation, particularly from coal-fired plants (7.4\% compared to 2020), following government measures introduced to mitigate the gas crisis \citep{MASE2024PNIEC}. After the crisis, the relative advantage shifted again in favour of gas-fired generation.

These dynamics highlight a key limitation of carbon pricing: its effectiveness in driving fuel switching critically depends on relative fuel prices. When gas prices rise sharply, as observed during the energy crisis, the switching price increases significantly, weakening the ability of the carbon price signal to incentivize cleaner generation. As a result, even relatively high CO$_2$ prices may fail to prevent a temporary return to more carbon-intensive technologies such as coal. This evidence suggests that carbon pricing should be complemented by mechanisms that preserve its effectiveness during periods of high fuel price. In particular, crisis episodes call for targeted and temporary interventions aimed at stabilizing relative price signals without undermining long-term decarbonization incentives.

\section{Concluding remarks and policy implications}\label{sec.conclusions}
This study investigates the impact of carbon pricing under the EU ETS on the Italian electricity market, with a particular focus on the CPTR across market zones during Phases 3 and 4 (2016-2024). The results show that carbon costs are positively and significantly transmitted to wholesale electricity prices, confirming the relevance of the EU ETS in influencing electricity market dynamics. However, the estimated CPTRs are consistently below 100\%, indicating incomplete pass-through. This suggests that electricity producers absorb part of the carbon costs rather than fully passing them on to consumers.\\
At the national level, the pass-through estimate is around 32\%, with no statistically significant change between Phase~3 and Phase~4. Substantial heterogeneity across Italian market zones emerges. While all zones exhibit positive CPTRs in Phase~3, their evolution in Phase~4 differs markedly. Northern and Centre-North zones, along with Sardinia, experience an increase in pass-through rates, whereas the Centre-South and Sicily show a decline. These differences can be due to variations in energy mix, carbon intensity, and infrastructure constraints, highlighting the importance of zonal characteristics within a unified carbon pricing framework. In addition, quantile regression results confirm that carbon cost pass-through is positive but incomplete across all market zones during Phase~3, with no strong dependence on price levels. For Phase~4, no significant change emerges at the aggregate level, although some zonal heterogeneity is observed, especially in Sardinia.

Findings are in line with those obtained by similar studies on carbon cost pass-through in the Italian electricity market. Although previous contributions rely on different methodologies, datasets, and model specifications, a common pattern emerges, namely a decline in the CPTR over time and across the phases of the EU ETS and market zones heterogeneity.\\
A first relevant contribution is provided by \cite{Mazzarella2017}, who documents a substantial reduction in the CPTR across the first three phases of the EU ETS. In particular, the estimated pass-through decreases from approximately 70\% in Phase 1 (2005--2007) to about 30\% in Phase 2 (2008--2012), and becomes close to zero in Phase~3 (2013--2015). This sharp decline is mainly attributed to changes in the supply side, including the evolving composition of market participants and the increasing role of financial operators.\\
Similarly, \cite{Caporin2021}, using a vector error correction model (VECM), find evidence of a relatively low carbon pass-through in the Italian electricity market in the long run. Their results suggest that, compared to other cost components such as natural gas, carbon costs play a more limited role in explaining electricity price dynamics, confirming the presence of incomplete and possibly weakening pass-through over time.\\
In addition to the temporal dimension, previous studies have also highlighted the importance of spatial heterogeneity within the Italian electricity market. In particular, \cite{CHERNYAVSKA2008} and \cite{Chernyavs2008}, adopting a load duration curve approach, estimate zonal CPTRs for different market areas, including the North and South zones during Phase 1. Their results suggest that pass-through rates vary significantly across the two zones in peak-period, reflecting differences in market structure, generation mix, and demand conditions.

This study could be extended in several directions. First, future research could strengthen the identification of causal relationship among variables by expanding the framework in Appendix \ref{sec.appendix.causality}.\\
Further insights could be obtained by introducing a higher level of temporal disaggregation. In particular, estimating the CPTR separately for different time intervals, such as peak and off-peak hours, would allow capturing potential intra-day heterogeneity in cost transmission \citep{Chernyavs2008, CHERNYAVSKA2008, SIJM2008, JOUVET20131370, Caporin2021}. This approach could shed light on how demand conditions and generation constraints influence the pass-through rate across different periods of the day.

\bibliographystyle{apalike}
\bibliography{References.bib}

\paragraph{Declaration of interests} The authors declare no conflicts of interest.

\paragraph{Acknowledgments} This study was funded by the European Union - NextGenerationEU, Mission 4, Component 2, in the framework of the GRINS - Growing Resilient, INclusive and Sustainable project (GRINS PE00000018 - CUP C93C22005270001). The views and opinions expressed are solely those of the authors and do not necessarily reflect those of the European Union, nor can the European Union be held responsible for them.

The authors gratefully acknowledges C. Wegener and M. Caporin for their valuable comments at the CFE-CMStatistics 2025 Conference held in London.

\paragraph{Data availability statement} 
The authors do not have permission to share the financial and electricity price data. To access the data, please contact the data providers cited in the paper.

\appendix

\section{Causality and cointegration issues}
\label{sec.appendix.causality}

\renewcommand{\thefigure}{A\arabic{figure}}
\renewcommand{\thetable}{A\arabic{table}}
\renewcommand{\theequation}{A\arabic{equation}}

\setcounter{figure}{0}
\setcounter{table}{0}

The baseline pass-through model estimates the elasticity of electricity prices with respect
to carbon costs. This section complements the CPTR estimates by testing whether carbon
costs contain predictive information for the dynamics of the electricity--fuel spread. The
analysis is based on the bivariate system $y_t = \left(s_t,\, C_t\right)'$, where \(s_t\) denotes the spread between electricity prices and fuel costs, and \(C_t\) denotes the carbon cost. Three complementary tests are implemented: standard Granger causality in stationary transformed differences, Toda--Yamamoto causality in levels, and VECM-based short-run, long-run, and joint causality \citep{Granger1969,TodaYamamoto1995,Johansen1991,EngleGranger1987}.

Let \(x_t\) denote the stationary transformed system. For the spread, the stationary
transformation follows the construction used in the empirical model:
\[
  \Delta \tilde{s}_t
  =
  \Delta \left(\log p^e_t - \log p^f_t\right),
\]
while for carbon costs the transformation is the log difference, \(\Delta \log C_t\). The
stationary vector is therefore $x_t =
  \left(
  \Delta \tilde{s}_t,\,
  \Delta \log C_t
  \right)'$. The unrestricted VAR(\(p\)) model is defined as
\begin{equation}
  x_t =
  c + A_1 x_{t-1} + A_2 x_{t-2} + \cdots + A_p x_{t-p} + u_t,
  \label{eq:var_differences_carbon_spread}
\end{equation}
where \(c\) is a vector of intercepts, \(A_k\) are coefficient matrices, and \(u_t\) is a
vector of innovations. The lag length \(p\) is selected by AIC with a maximum of 21 lags. The spread equation implied by Eq.~\eqref{eq:var_differences_carbon_spread} is
\begin{equation}
  \Delta \tilde{s}_t =
  c_s
  + \sum_{k=1}^{p} A_{ss,k}\Delta \tilde{s}_{t-k}
  + \sum_{k=1}^{p} A_{sC,k}\Delta \log C_{t-k}
  + u_{s,t}.
  \label{eq:granger_spread_equation}
\end{equation}
The standard Granger-causality hypotheses from carbon costs to the spread are
\begin{align}
  H_{0}^{G}:&
  \quad
  A_{sC,1}=A_{sC,2}=\cdots=A_{sC,p}=0,
  \label{eq:hyp_granger_carbon_spread_null}
  \\
  H_{1}^{G}:&
  \quad
  \exists k \in \{1,\ldots,p\}
  \text{ such that } A_{sC,k}\neq 0.
  \label{eq:hyp_granger_carbon_spread_alt}
\end{align}
The null hypothesis is tested using a Wald exclusion statistic. Let
\(\widehat{\theta}^{G}_{C}\) be the vector of estimated coefficients on the \(p\) lags of
carbon costs in the spread equation, and let
\(\widehat{V}(\widehat{\theta}^{G}_{C})\) be the corresponding estimated covariance
matrix. The Wald statistic is
\begin{equation}
  W_{C}^{G}
  =
  \widehat{\theta}^{G\prime}_{C}
  \left[
  \widehat{V}(\widehat{\theta}^{G}_{C})
  \right]^{-1}
  \widehat{\theta}^{G}_{C}.
  \label{eq:granger_wald_carbon_spread}
\end{equation}
Under the null hypothesis, \(W_{C}^{G}\) is asymptotically distributed as
\(\chi^2(p)\). Rejection of \(H_{0}^{G}\) implies that past changes in carbon costs help
predict changes in the electricity--fuel spread, after controlling for lagged spread
dynamics.

The Toda--Yamamoto test is used to check that the causality results are not driven by pre-testing choices on stationarity and cointegration. Instead of estimating the model in differences, the procedure estimates an augmented VAR in levels:
\begin{equation}
  y_t =
  a + B_1 y_{t-1} + B_2 y_{t-2}
  + \cdots + B_{p+d_{\max}} y_{t-p-d_{\max}} + e_t,
  \label{eq:ty_augmented_var_carbon_spread}
\end{equation}
where \(p\) is the AIC-selected lag length and \(d_{\max}=1\) is the maximum order of
integration allowed by the unit-root results. The additional \(d_{\max}\) lag is included
to guarantee valid asymptotic inference, but it is not included in the Wald restrictions.

The spread equation of the augmented VAR is
\begin{equation}
  s_t =
  a_s
  + \sum_{k=1}^{p+d_{\max}} B_{ss,k}s_{t-k}
  + \sum_{k=1}^{p+d_{\max}} B_{sC,k}C_{t-k}
  + e_{s,t}.
  \label{eq:ty_spread_equation}
\end{equation}
The Toda--Yamamoto hypotheses from carbon costs to the spread are
\begin{align}
  H_{0}^{TY}:&
  \quad
  B_{sC,1}=B_{sC,2}=\cdots=B_{sC,p}=0,
  \label{eq:hyp_ty_carbon_spread_null}
  \\
  H_{1}^{TY}:&
  \quad
  \exists k \in \{1,\ldots,p\}
  \text{ such that } B_{sC,k}\neq 0.
  \label{eq:hyp_ty_carbon_spread_alt}
\end{align}
The restrictions are imposed only on the first \(p\) lags, while the additional
\(d_{\max}\) lag is included in the augmented VAR but excluded from the test. Let
\(\widehat{\theta}^{TY}_{C}\) denote the vector of estimated coefficients on the first
\(p\) lags of carbon costs in the spread equation. The Toda--Yamamoto modified Wald
statistic is
\begin{equation}
  W_{C}^{TY}
  =
  \widehat{\theta}^{TY\prime}_{C}
  \left[
  \widehat{V}\left(\widehat{\theta}^{TY}_{C}\right)
  \right]^{-1}
  \widehat{\theta}^{TY}_{C}.
  \label{eq:ty_mwald_carbon_spread}
\end{equation}
Under the null hypothesis, \(W_{C}^{TY}\) is asymptotically distributed as
\(\chi^2(p)\). Rejection of \(H_{0}^{TY}\) provides evidence of predictive causality from
carbon costs to the electricity--fuel spread that is robust to the integration and
cointegration properties of the variables.

In the presence of cointegration, the system can be represented as a cointegrated VAR.
Starting from a VAR(\(K\)) in levels,
\begin{equation}
  y_t =
  \mu + \Pi_1 y_{t-1} + \Pi_2 y_{t-2}
  + \cdots + \Pi_K y_{t-K} + \varepsilon_t,
  \label{eq:level_var_for_vecm_carbon_spread}
\end{equation}
the corresponding VECM representation is
\begin{equation}
  \Delta y_t =
  \mu
  + \Pi y_{t-1}
  + \sum_{k=1}^{K-1}\Gamma_k \Delta y_{t-k}
  + \varepsilon_t,
  \label{eq:vecm_general_carbon_spread}
\end{equation}
where
\[
  \Pi = -\left(I-\Pi_1-\Pi_2-\cdots-\Pi_K\right),
  \qquad
  \Gamma_k = -\left(\Pi_{k+1}+\cdots+\Pi_K\right).
\]
If the matrix \(\Pi\) has reduced rank \(r<n\), it can be decomposed as $\Pi = \alpha \beta'$, where \(\beta\) contains the \(r\) cointegrating vectors and \(\alpha\) contains the
adjustment coefficients. The estimated VECM is therefore
\begin{equation}
  \Delta y_t =
  \mu
  + \alpha \beta' y_{t-1}
  + \sum_{k=1}^{K-1}\Gamma_k \Delta y_{t-k}
  + \varepsilon_t.
  \label{eq:vecm_robustness_carbon_spread}
\end{equation}
The term \(\beta' y_{t-1}\) contains the \(r\) error-correction terms, while \(\alpha\)
measures the speed at which each variable adjusts to deviations from the long-run
equilibria. The Johansen system is estimated with a constant in the cointegration space
and the transitory specification.

The spread equation in the VECM can be written as
\begin{equation}
  \Delta s_t =
  \mu_s
  + \sum_{m=1}^{r} \alpha_{s,m}ECT_{m,t-1}
  + \sum_{k=1}^{K-1}\Gamma_{ss,k}\Delta s_{t-k}
  + \sum_{k=1}^{K-1}\Gamma_{sC,k}\Delta C_{t-k}
  + \varepsilon_{s,t},
  \label{eq:vecm_spread_equation}
\end{equation}
where \(ECT_{m,t-1}\) denotes the \(m\)-th error-correction term.

Within the VECM, three forms of causality are tested. Short-run causality from carbon
costs to the spread is assessed by testing whether the coefficients on lagged first
differences of carbon costs are jointly zero:
\begin{align}
  H_{0}^{SR}:&
  \quad
  \Gamma_{sC,1}=\Gamma_{sC,2}=\cdots=\Gamma_{sC,K-1}=0,
  \label{eq:hyp_vecm_sr_carbon_spread_null}
  \\
  H_{1}^{SR}:&
  \quad
  \exists k \in \{1,\ldots,K-1\}
  \text{ such that } \Gamma_{sC,k}\neq 0.
  \label{eq:hyp_vecm_sr_carbon_spread_alt}
\end{align}
Let \(\widehat{\theta}^{SR}_{C}\) denote the vector of estimated short-run coefficients
associated with carbon costs in the spread equation:
\[
  \widehat{\theta}^{SR}_{C}
  =
  \left(
  \widehat{\Gamma}_{sC,1},
  \widehat{\Gamma}_{sC,2},
  \ldots,
  \widehat{\Gamma}_{sC,K-1}
  \right)'.
\]
The short-run Wald statistic is
\begin{equation}
  W_{C}^{SR}
  =
  \widehat{\theta}^{SR\prime}_{C}
  \left[
  \widehat{V}\left(\widehat{\theta}^{SR}_{C}\right)
  \right]^{-1}
  \widehat{\theta}^{SR}_{C}.
  \label{eq:vecm_sr_wald_carbon_spread}
\end{equation}
Under the null hypothesis, \(W_{C}^{SR} \sim \chi^2(K-1)\). Rejection of
\(H_{0}^{SR}\) indicates short-run predictive causality from carbon costs to the
electricity--fuel spread.

Long-run causality is tested through the joint significance of the adjustment coefficients
associated with the error-correction terms in the spread equation:
\begin{align}
  H_{0}^{LR}:&
  \quad
  \alpha_{s,1}=\alpha_{s,2}=\cdots=\alpha_{s,r}=0,
  \label{eq:hyp_vecm_lr_carbon_spread_null}
  \\
  H_{1}^{LR}:&
  \quad
  \exists m \in \{1,\ldots,r\}
  \text{ such that } \alpha_{s,m}\neq 0.
  \label{eq:hyp_vecm_lr_carbon_spread_alt}
\end{align}
Let \(\widehat{\theta}^{LR}\) denote the vector of estimated adjustment coefficients in
the spread equation:
\[
  \widehat{\theta}^{LR}
  =
  \left(
  \widehat{\alpha}_{s,1},
  \widehat{\alpha}_{s,2},
  \ldots,
  \widehat{\alpha}_{s,r}
  \right)'.
\]
The long-run Wald statistic is
\begin{equation}
  W^{LR}
  =
  \widehat{\theta}^{LR\prime}
  \left[
  \widehat{V}\left(\widehat{\theta}^{LR}\right)
  \right]^{-1}
  \widehat{\theta}^{LR}.
  \label{eq:vecm_lr_wald_carbon_spread}
\end{equation}
Under the null hypothesis, \(W^{LR} \sim \chi^2(r)\). Rejection of \(H_{0}^{LR}\)
implies that the electricity--fuel spread adjusts to deviations from the long-run
equilibrium linking the spread and carbon costs. Joint causality combines the short-run restrictions on carbon costs and the long-run
error-correction restrictions:
\begin{align}
  H_{0}^{JC}:&
  \quad
  \Gamma_{sC,1}=\cdots=\Gamma_{sC,K-1}=0
  \quad \text{and} \quad
  \alpha_{s,1}=\cdots=\alpha_{s,r}=0,
  \label{eq:hyp_vecm_jc_carbon_spread_null}
  \\
  H_{1}^{JC}:&
  \quad
  \exists k \text{ such that } \Gamma_{sC,k}\neq 0
  \quad \text{or} \quad
  \exists m \text{ such that } \alpha_{s,m}\neq 0.
  \label{eq:hyp_vecm_jc_carbon_spread_alt}
\end{align}
Let
\[
  \widehat{\theta}^{JC}_{C}
  =
  \left(
  \widehat{\Gamma}_{sC,1},
  \ldots,
  \widehat{\Gamma}_{sC,K-1},
  \widehat{\alpha}_{s,1},
  \ldots,
  \widehat{\alpha}_{s,r}
  \right)'.
\]
The joint-causality Wald statistic is
\begin{equation}
  W_{C}^{JC}
  =
  \widehat{\theta}^{JC\prime}_{C}
  \left[
  \widehat{V}\left(\widehat{\theta}^{JC}_{C}\right)
  \right]^{-1}
  \widehat{\theta}^{JC}_{C}.
  \label{eq:vecm_jc_wald_carbon_spread}
\end{equation}
Under the null hypothesis, \(W_{C}^{JC} \sim \chi^2(K-1+r)\). Rejection of \(H_{0}^{JC}\) indicates that carbon costs contribute to spread dynamics either through short-run predictive effects, through long-run equilibrium correction, or through both
channels.

Table~\ref{tab:carbon_causality_spread} reports the causality-test results from carbon costs to the electricity--fuel spread. The standard Granger tests reject the null of no
predictive content for Italy, North, Centre-North, and Sardinia at the 1\%, and for Centre-South at the 5\% and for Sicily at the 10\%. The Toda--Yamamoto tests reject the null in all zones, indicating that carbon costs contain predictive information for the electricity--fuel spread when the system is estimated in levels and inference is made robust to integration and cointegration properties.

The VECM-based tests provide evidence of short-run causality from carbon costs to the
spread in all zones for which a cointegrating relation is detected. Long-run adjustment is
also statistically significant in all cointegrated zones. Joint causality is detected in all
zones for which the VECM is estimated. For Sardinia, the Johansen procedure does not
select a positive cointegration rank at the 5 percent level, so the VECM-based short-run,
long-run, and joint causality statistics are not reported. Overall, the results suggest that carbon costs have predictive content for the electricity--fuel spread at the national level and across most market zones, supporting the existence of a causal relationship.

\begin{table}[H]
\centering
\begin{tabular}{lccccc}
\hline
Zone & $W^G$ & $W^{TY}$ & $W^{SR}$ & $W^{LR}$ & $W^{JC}$ \\
\hline
Italy & 48.62** & 81.63** & 88.78** & 13.41** & 98.20** \\
North & 109.90** & 172.30** & 186.23** & 13.29** & 194.91** \\
Centre-North & 57.76** & 84.14** & 79.65** & 13.28*** & 90.31** \\
Centre-South & 37.75* & 124.14** & 144.15** & 17.81** & 152.12** \\
Sicily & 28.69 & 59.59** & 61.50** & 8.72** & 68.42** \\
Sardinia & 53.58** & 48.16** & -- & -- & -- \\
\hline
\end{tabular}
\caption{Carbon costs and electricity--fuel spread: test statistics. Significance levels:** $p$-value $\leq 0.01$, * $p$-value $\leq 0.05$.}
\label{tab:carbon_causality_spread}
\end{table}

\renewcommand{\thefigure}{B\arabic{figure}}

\section{Figures}\label{sec.appendix.fig}
\setcounter{figure}{0}

\begin{figure}[H]
    \centering
    \includegraphics[width=0.98\textwidth]{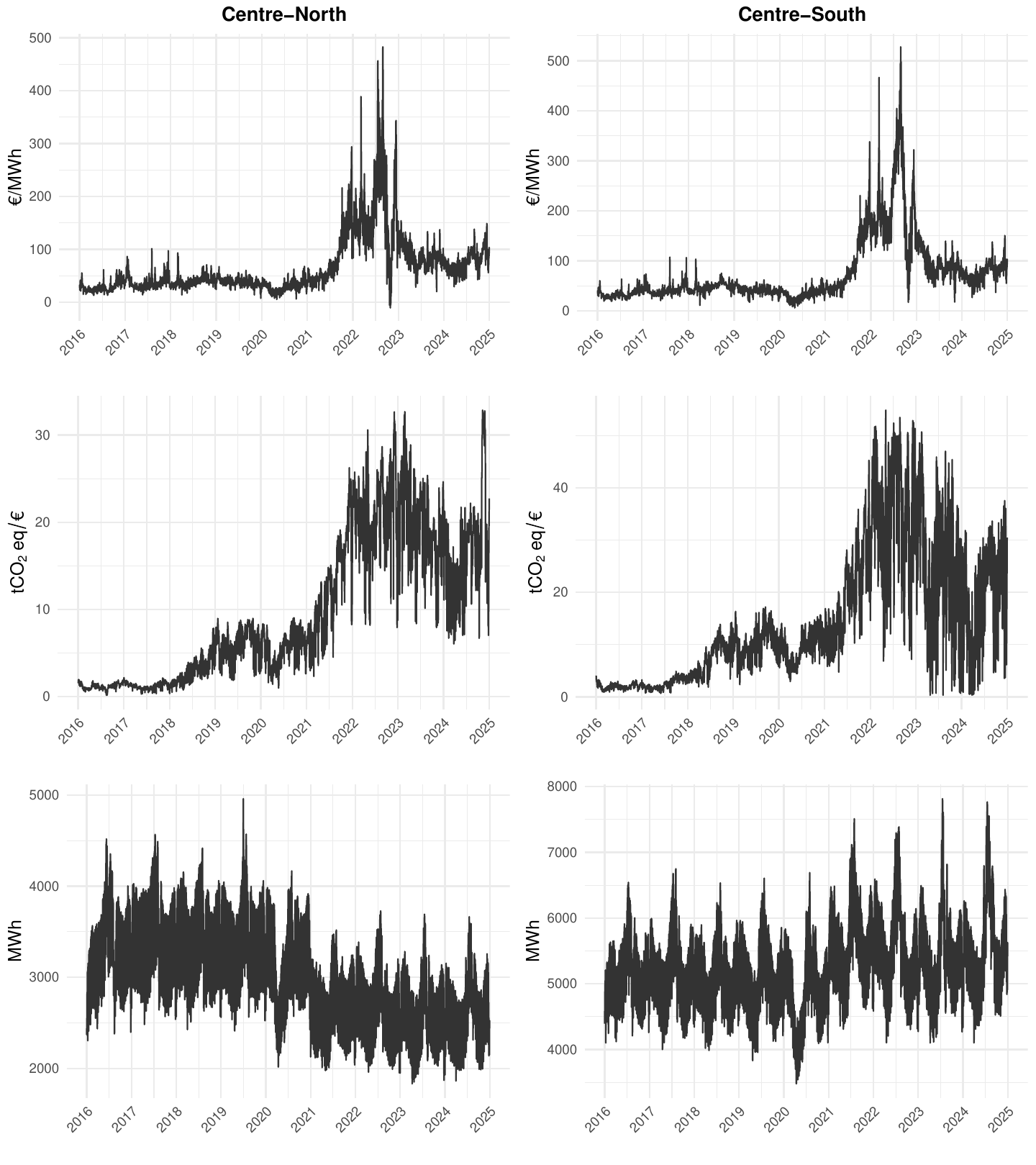}
    \caption{Daily spreads between electricity prices and fuel costs (top panels), carbon costs (middle panels), and electricity demand (bottom panels) for the Centre-North and Centre-South zones, from 1 January 2016 to 31 December 2024.}
    \label{fig:series.cnorth.csouth}
\end{figure}

\begin{figure}[H]
    \centering
    \includegraphics[width=0.98\textwidth]{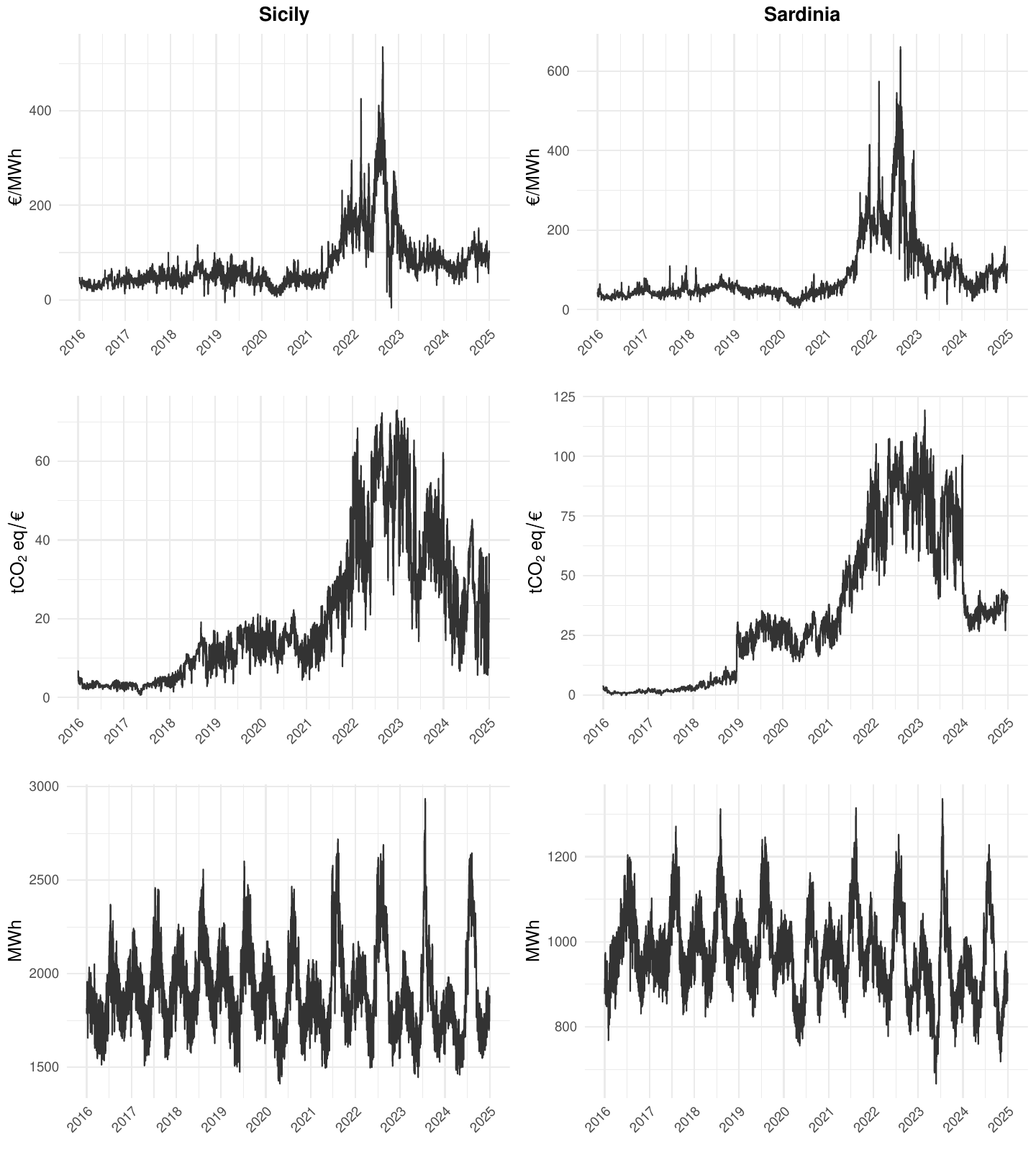}
    \caption{Daily spreads between electricity prices and fuel costs (top panels), carbon costs (middle panels), and electricity demand (bottom panels) for Sicily and Sardinia, from 1 January 2016 to 31 December 2024.}
    \label{fig:series.sicily.sardinia}
\end{figure}

\begin{figure}[H]
    \centering
    \includegraphics[width=1\linewidth]{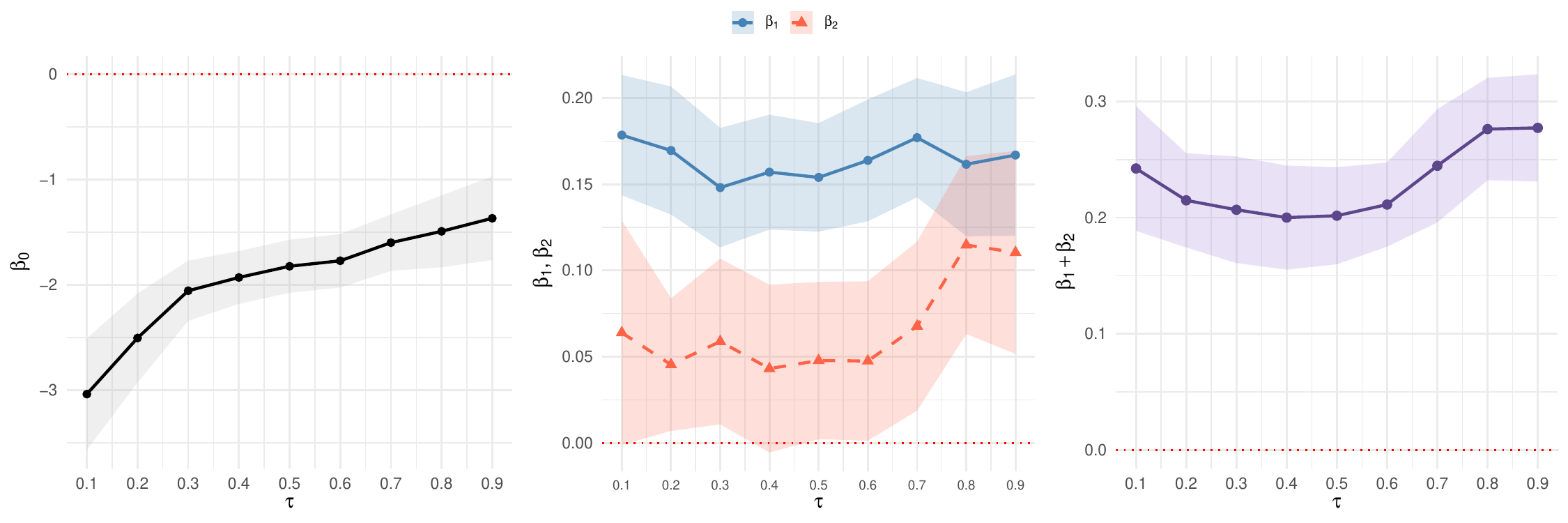}
    \caption{Quantile regression estimates for the North zone. The estimated coefficients $\beta_0$ (left panel), $\beta_1$ and $\beta_2$ (middle panel), and their sum $\beta_1 + \beta_2$ (right panel) are plotted across quantiles $\tau$. Shaded areas indicate $90\%$ confidence intervals.}
    \label{fig.quant.north}
\end{figure}

\begin{figure}[H]
    \centering
    \includegraphics[width=1\linewidth]{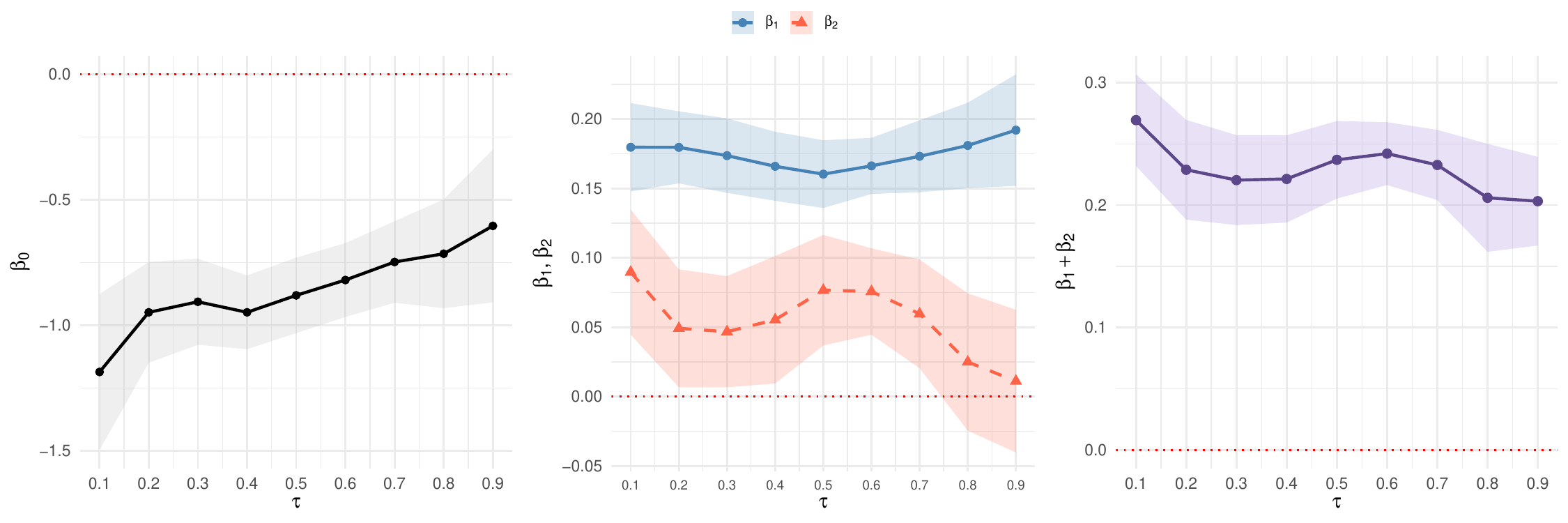}
    \caption{Quantile regression estimates for the Centre-North zone. The estimated coefficients $\beta_0$ (left panel), $\beta_1$ and $\beta_2$ (middle panel), and their sum $\beta_1 + \beta_2$ (right panel) are plotted across quantiles $\tau$. Shaded areas indicate $90\%$ confidence intervals.}
    \label{fig.quant.cnorth}
\end{figure}

\begin{figure}[H]
    \centering
    \includegraphics[width=1\linewidth]{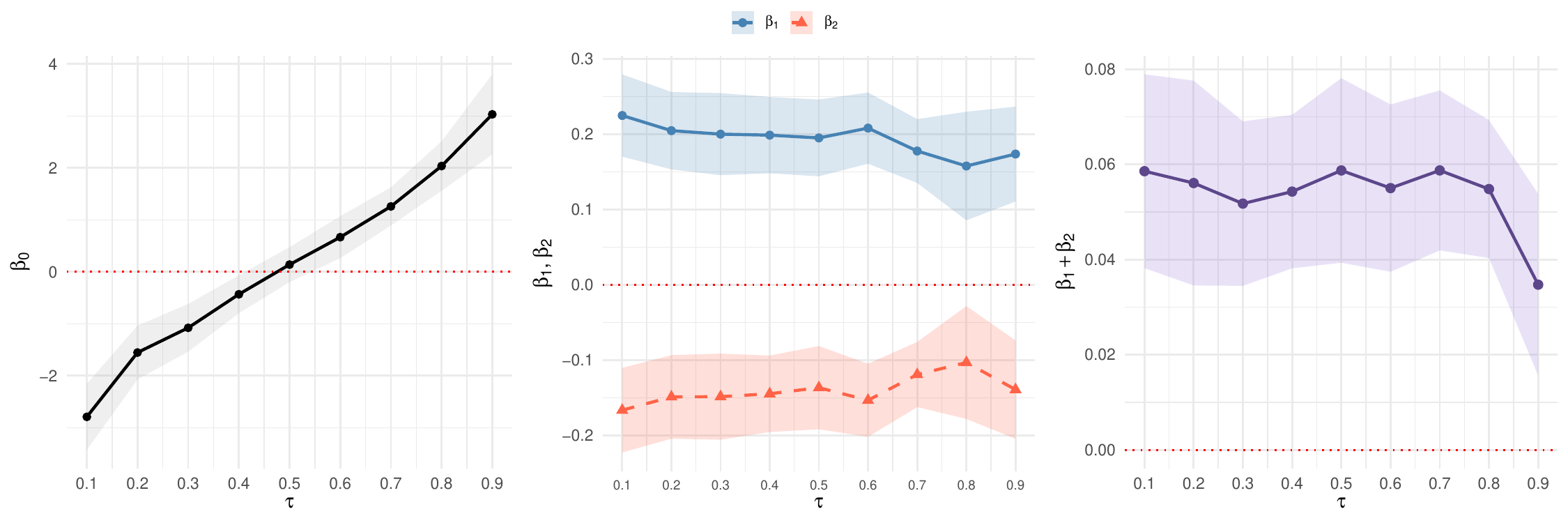}
    \caption{Quantile regression estimates for the Centre-South zone. The estimated coefficients $\beta_0$ (left panel), $\beta_1$ and $\beta_2$ (middle panel), and their sum $\beta_1 + \beta_2$ (right panel) are plotted across quantiles $\tau$. Shaded areas indicate $90\%$ confidence intervals.}
    \label{fig.quant.csouth}
\end{figure}

\begin{figure}[H]
    \centering
    \includegraphics[width=1\linewidth]{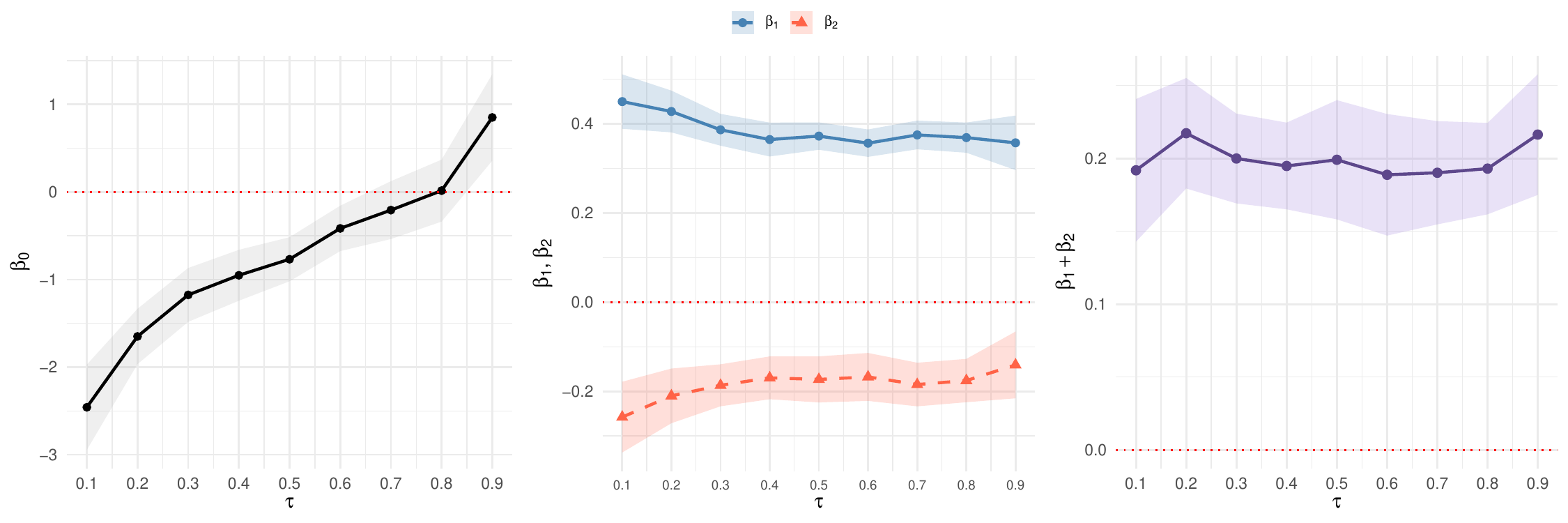}
    \caption{Quantile regression estimates for Sicily. The estimated coefficients $\beta_0$ (left panel), $\beta_1$ and $\beta_2$ (middle panel), and their sum $\beta_1 + \beta_2$ (right panel) are plotted across quantiles $\tau$. Shaded areas indicate $90\%$ confidence intervals.}
    \label{fig.quant.sicily}
\end{figure}

\begin{figure}[H]
    \centering
    \includegraphics[width=1\linewidth]{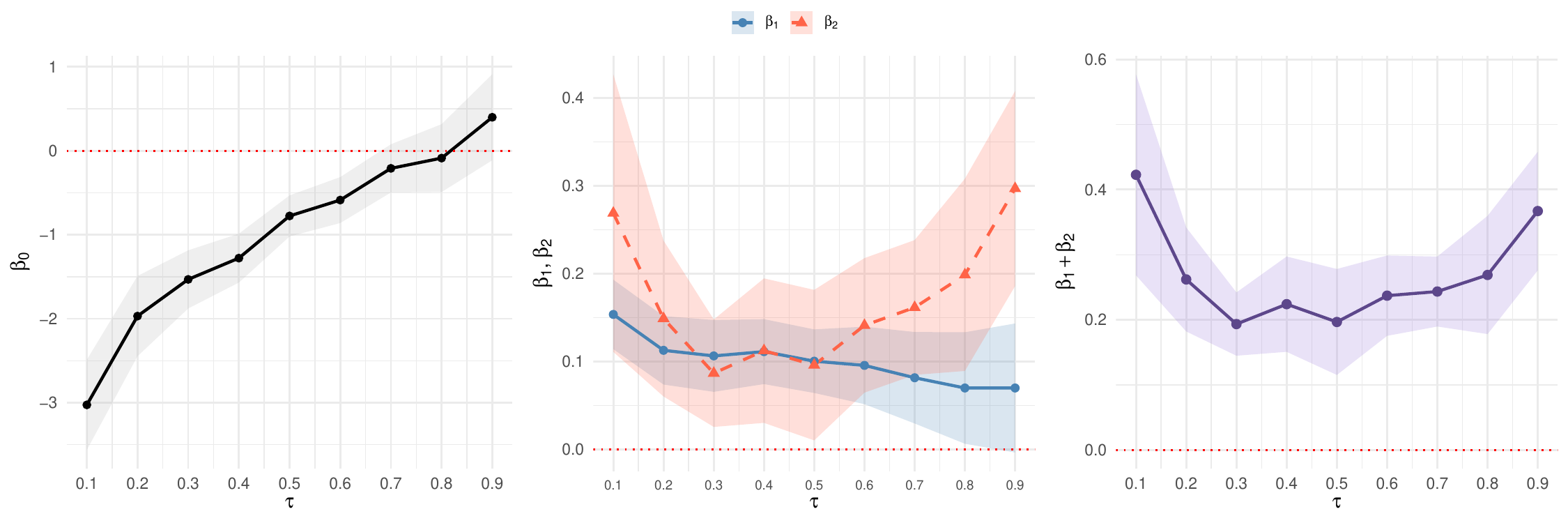}
    \caption{Quantile regression estimates for Sardinia. The estimated coefficients $\beta_0$ (left panel), $\beta_1$ and $\beta_2$ (middle panel), and their sum $\beta_1 + \beta_2$ (right panel) are plotted across quantiles $\tau$. Shaded areas indicate $90\%$ confidence intervals.}
    \label{fig.quant.sardinia}
\end{figure}

\renewcommand{\thetable}{C\arabic{table}}
\setcounter{table}{0}
\section{Tables}\label{sec.appendix.tab}

\begin{table}[ht]
\centering
\begin{tabular}{lrrrrrr}
  \hline
Variable & Italy & North & Centre-North & Centre-South & Sicily & Sardinia \\ 
  \hline
$\phi_1$    & 1.31 & 1.42 & 1.30 & 1.22 & 1.22 & 1.22 \\ 
$\phi_2$    & 1.43 & 1.53 & 1.43 & 1.39 & 1.36 & 1.35 \\ 
$\phi_3$    & 1.42 & 1.44 & 1.41 & 1.46 & 1.38 & 1.40 \\ 
$\phi_4$    & 1.34 & 1.33 & 1.31 & 1.39 & 1.35 & 1.34 \\ 
$\phi_5$    & 1.27 & 1.35 & 1.32 & 1.26 & 1.21 & 1.24 \\ 
$\phi_7$    & 1.21 & 1.48 & 1.40 & 1.10 & 1.10 & 1.12 \\ 
$\phi_{14}$ & 1.19 & 1.45 & 1.36 & 1.05 & 1.07 & 1.08 \\ 
$\phi_{21}$ & 1.18 & 1.43 & 1.35 & 1.05 & 1.06 & 1.07 \\ 
  \hline
\end{tabular}
\caption{Variance inflation factors for the autoregressive terms in the baseline specification.}
\label{tab:vif_baseline}
\end{table}

\begin{table}[H]
\begin{center}
\begin{tabular}{lcccccc}
\hline
\multicolumn{1}{l}{\rule{0pt}{12pt}}
 & Italy & North & Centre--North & Centre--South & Sicily & Sardinia \\[2pt]
\hline\rule{0pt}{12pt}$\beta_0$ & -1.87** & -2.02** & -0.93** & 0.001 & -1.01** & -1.16** \\
$\phi_1$ & -0.34** & -0.33** & -0.27** & -0.38** & -0.31** & -0.33** \\
$\phi_2$ & -0.27** & -0.27** & -0.23** & -0.34** & -0.22** & -0.29** \\
$\phi_3$ & -0.21** & -0.23** & -0.18** & -0.25** & -0.17** & -0.22** \\
$\phi_4$ & -0.15** & -0.16** & -0.13** & -0.21** & -0.13** & -0.16** \\
$\phi_5$ & -0.12** & -0.14** & -0.13** & -0.18** & -0.10** & -0.11** \\
$\phi_7$ & 0.03 & 0.07** & 0.09** & 0.06** & 0.05** & 0.08** \\
$\phi_{14}$ & 0.03 & 0.06** & 0.08** & 0.09** & 0.08** & 0.10** \\
$\phi_{21}$ & 0.08** & 0.10** & 0.11** & 0.09** & 0.10** & 0.13** \\
$\beta_1$ & 0.32** & 0.16** & 0.18** & 0.20** & 0.41** & 0.12** \\
$\beta_2$ & -0.04 & 0.07* & 0.06* & -0.16** & -0.21** & 0.29** \\
$\beta_3$ & 0.18** & 0.21** & 0.12** & -0.0002 & 0.13** & 0.17** \\
\hline
$R_{Adj}^2$ & 37.31 & 43.81 & 44.13 & 25.10 & 33.30 & 25.34 \\[2pt]
\hline
\end{tabular}
\end{center}
\caption{Estimated carbon pass-through rates and regression results using average daily prices. Significance levels: ** p-value$\leq0.01$, * p-value$\leq0.05$.}
\label{tab:cptr_estimates_dailyavg}
\end{table}

\begin{table}[H]
\begin{center}
\begin{tabular}{lcccccc}
\hline
\multicolumn{1}{l}{\rule{0pt}{12pt}}
 & Italy & North & Centre--North & Centre--South & Sicily & Sardinia \\[2pt]
\hline\rule{0pt}{12pt}$\beta_0$ & -41.95** & -7.67 & -7.35* & -51.40** & -50.46** & -16.81 \\
$\phi_1$ & -0.38** & -0.37** & -0.31** & -0.39** & -0.34** & -0.35** \\
$\phi_2$ & -0.28** & -0.28** & -0.25** & -0.33** & -0.22** & -0.29** \\
$\phi_3$ & -0.22** & -0.23** & -0.19** & -0.26** & -0.18** & -0.22** \\
$\phi_4$ & -0.16** & -0.17** & -0.14** & -0.21** & -0.13** & -0.16** \\
$\phi_5$ & -0.13** & -0.14** & -0.14** & -0.17** & -0.10** & -0.11** \\
$\phi_7$ & 0.04* & 0.08** & 0.10** & 0.05* & 0.05** & 0.08** \\
$\phi_{14}$ & 0.05* & 0.07** & 0.09** & 0.08** & 0.08** & 0.10** \\
$\phi_{21}$ & 0.09** & 0.11** & 0.12** & 0.08** & 0.10** & 0.13** \\
$\beta_1$ & 0.32** & 0.17** & 0.18** & 0.20** & 0.40** & 0.12** \\
$\beta_2$ & -0.04 & 0.07* & 0.05* & -0.15** & -0.21** & 0.29** \\
$\beta_3$ & 7.95** & 1.37 & 1.72 & 11.99** & 13.18** & 4.72 \\
$\beta_4$ & -0.38** & -0.06 & -0.10 & -0.70** & -0.86** & -0.33 \\
\hline
$R_{Adj}^2$ & 39.90 & 46.58 & 46.20 & 25.91 & 35.56 & 26.37 \\[2pt]
\hline
\end{tabular}
\end{center}
\caption{Estimated carbon pass-through rates and regression results for the model with quadratic demand. The coefficient $\beta_4$ is associated to the quadratic term $\log(D_t)^2$. Significance levels: ** p-value$\leq0.01$, * p-value$\leq0.05$.}
\label{tab:cptr_estimates_quad}
\end{table}

\begin{table}[H]
\begin{center}
\begin{tabular}{lcccccc}
\hline
\multicolumn{1}{l}{\rule{0pt}{12pt}}
 & Italy & North & Centre--North & Centre--South & Sicily & Sardinia \\[2pt]
\hline\rule{0pt}{12pt}$\beta_0$ 
& $<0.01$ & $<0.01$ & $<0.01$ & $<0.01$ & $<0.01$ & $<0.01$ \\

$\phi_1$ 
& -0.38** & -0.37** & -0.31** & -0.39** & -0.34** & -0.35** \\

$\phi_2$ 
& -0.28** & -0.28** & -0.25** & -0.33** & -0.22** & -0.29** \\

$\phi_3$ 
& -0.22** & -0.23** & -0.19** & -0.25** & -0.18** & -0.22** \\

$\phi_4$ 
& -0.16** & -0.16** & -0.14** & -0.21** & -0.13** & -0.16** \\

$\phi_5$ 
& -0.13** & -0.14** & -0.14** & -0.17** & -0.10** & -0.11** \\

$\phi_7$ 
& 0.04** & 0.08** & 0.10** & 0.05** & 0.05** & 0.08** \\

$\phi_{14}$ 
& 0.05** & 0.07** & 0.09** & 0.08** & 0.08** & 0.10** \\

$\phi_{21}$ 
& 0.09** & 0.11** & 0.12** & 0.08** & 0.10** & 0.13** \\

$\beta_1$ 
& 0.32** & 0.17** & 0.18** & 0.20** & 0.41** & 0.12** \\

$\beta_2$ 
& -0.04 & 0.07** & 0.05** & -0.15** & -0.21** & 0.29** \\

edf 
& 1.96** & 1.01** & 1.79** & 1.95** & 1.97** & 1.68** \\[2pt]

\hline
$R^2_{Adj}$ 
& 39.89 & 46.59 & 46.21 & 25.83 & 35.49 & 26.37 \\[2pt]
\hline
\end{tabular}
\end{center}
\caption{Estimated carbon pass-through rates and regression results for the GAM specification. Effective degrees of freedom (edf) are reported for the smoothing term. Significance levels: ** p-value$\leq0.01$.}
\label{tab:cptr_estimates_gam}
\end{table}

\begin{table}[H]
\centering
\begin{tabular}{lcccccc}
  \hline
 & Italy & North & Centre--North & Centre--South & Sicily & Sardinia \\[2pt]
  \hline
$\beta_0$ & -1.86** & -2.03** & -0.99** & 0.03 & -1.02** & -1.19** \\ 
  $\phi_1$ & -0.38** & -0.37** & -0.31** & -0.39** & -0.34** & -0.35** \\ 
  $\phi_2$ & -0.28** & -0.28** & -0.25** & -0.33** & -0.22** & -0.29** \\ 
  $\phi_3$ & -0.22** & -0.23** & -0.19** & -0.25** & -0.18** & -0.22** \\ 
  $\phi_4$ & -0.16** & -0.16** & -0.14** & -0.21** & -0.13** & -0.16** \\ 
  $\phi_5$ & -0.12** & -0.14** & -0.14** & -0.18** & -0.10** & -0.11** \\ 
  $\phi_7$ & 0.04* & 0.08** & 0.10** & 0.05** & 0.05** & 0.08** \\ 
  $\phi_{14}$ & 0.05* & 0.07** & 0.09** & 0.08** & 0.09** & 0.10** \\ 
  $\phi_{21}$ & 0.09** & 0.11** & 0.12** & 0.08** & 0.10** & 0.13** \\ 
  $\beta_1$ & 0.32** & 0.17** & 0.18** & 0.20** & 0.41** & 0.12** \\ 
  $\beta_2$ & -0.03 & 0.07* & 0.06** & -0.15** & -0.21** & 0.29** \\ 
  $\beta_3$ & 0.18** & 0.21** & 0.12** & < 0.01 & 0.14** & 0.17** \\ 
  $\beta_4$ & 0.01 & $<0.01$ & 0.02* & 0.01 & 0.01 & 0.01 \\ 
  \hline

  $R^2_{Adj}$ & 39.49 & 46.57 & 46.24 & 25.40 & 34.84 & 26.30 \\[2pt]
   \hline
\end{tabular}
\caption{Estimated carbon pass-through rates and regression results for the model with the coal-maximization programme dummy. The coefficient $\beta_4$ is associated with the dummy equal to one from September 2022 to March 2023. Significance levels: ** p-value$\leq$ 0.01, * p-value$\leq$ 0.05.} 
\label{tab:coal_max_dummy}
\end{table}

\begin{table}[H]
\centering
\begin{tabular}{lcccccc}
  \hline
& Italy & North & Centre--North & Centre--South & Sicily & Sardinia \\ [2pt]
  \hline
$\beta_0$ & -1.34** & -1.34** & -0.23* & -0.35 & -0.90** & -0.78** \\ 
  $\phi_1$ & -0.38** & -0.38** & -0.35** & -0.40** & -0.34** & -0.37** \\ 
  $\phi_2$ & -0.28** & -0.28** & -0.26** & -0.34** & -0.21** & -0.28** \\ 
  $\phi_3$ & -0.22** & -0.22** & -0.19** & -0.25** & -0.19** & -0.21** \\ 
  $\phi_4$ & -0.16** & -0.15** & -0.13** & -0.21** & -0.14** & -0.14** \\ 
  $\phi_5$ & -0.12** & -0.12** & -0.10** & -0.17** & -0.09** & -0.07** \\ 
  $\phi_7$ & 0.01 & 0.03 & 0.02 & 0.05* & < 0.01 & < 0.01 \\ 
  $\phi_{14}$ & 0.02 & 0.02 & 0.01 & 0.08** & 0.03 & 0.02 \\ 
  $\phi_{21}$ & 0.06** & 0.06** & 0.04 & 0.08** & 0.05* & 0.05** \\ 
  $\beta_1$ & 0.29** & 0.13** & 0.14** & 0.24** & 0.41** & 0.11** \\ 
  $\beta_2$ & -0.03 & 0.03 & 0.05* & -0.17** & -0.22** & 0.29** \\ 
  $\beta_3$ & 0.13** & 0.14** & 0.04** & 0.03 & 0.13** & 0.13** \\ 
  \hline
  $R^2_{Adj}$ & 40.73 & 48.94 & 50.31 & 26.44 & 39.07 & 32.79 \\[2pt]
   \hline
\end{tabular}
\caption{Estimated carbon pass-through rates and regression results for the model with day-of-week and month fixed effects. Fixed-effect coefficients are included in the regression but not reported. Significance levels: ** p-value$\leq$ 0.01, * p-value$\leq$ 0.05.} 
\label{tab:week_month_fixed_effects}
\end{table}

\end{document}